\documentclass[prb,aps,nofootinbib,preprintnumbers,superscriptaddress,floatfix,twocolumn]{revtex4-2}
\usepackage{graphicx}
\usepackage{color}
\usepackage{bm}
\usepackage{amsmath}
\usepackage{amssymb}
\usepackage{float}
\usepackage{dblfloatfix}
\usepackage{physics}
\usepackage{xspace}
\usepackage[normalem]{ulem}
\usepackage{units}
\usepackage{dcolumn}
\usepackage{dblfloatfix} 
\usepackage{paralist}
\usepackage[percent]{overpic} 
\usepackage{natmove}
\usepackage{tikz}
\usepackage{natbib}
\usepackage{array}
\usepackage{xcolor}
\usepackage{comment}
\usepackage[table]{xcolor}
\usepackage{tikz-3dplot}
\newcommand{\blue}[1]{{\color{blue}#1}}
\newcommand{\red}[1]{{\color{red}#1}}

\usepackage{hyperref}
\hypersetup{
    colorlinks=true,
    linkcolor=blue,      
    urlcolor=cyan,       
    citecolor=cyan      
}

\begin{document}

    \title{A Mean-Field Lindblad Master Equation Framework for Interaction-Driven Decoherence in Solid-State Qubit Ensembles}
    \author{Dhiman~Nandi}
    \affiliation{Department of Chemistry, University of Colorado Boulder, Boulder, Colorado 80309, USA}
    \affiliation{Ann and H.J. Smead Aerospace Engineering Sciences, University of Colorado Boulder, Boulder, CO 80303, USA}
    \author{Sanghamitra~Neogi}
    \email{sanghamitra.neogi@colorado.edu}
    \affiliation{Ann and H.J. Smead Aerospace Engineering Sciences, University of Colorado Boulder, Boulder, CO 80303, USA}
        

\begin{abstract}
Multi-qubit systems are essential for scalable quantum technologies, but their performance is often limited by decoherence arising from qubit--qubit interactions and environmental noise. Although previous studies have investigated environmental decoherence in single-qubit systems, as well as gate fidelity and entanglement measures in multi-qubit systems, a predictive framework that connects qubit--qubit interactions, concentration, spatial distribution, and bath occupation to relaxation and decoherence times remains lacking. Here, we develop a multi-qubit mean-field Lindblad master equation (MQMF-LME) framework to describe the population and coherence dynamics of a solid-state qubit embedded in an interacting multi-qubit environment. The framework treats one qubit as the system of interest and the surrounding qubits as an effective bath, incorporating intrinsic relaxation, excitation transfer from the system to the bath, and excitation transfer from thermally populated bath qubits back to the system. Analytical solutions provide closed-form expressions for the density-matrix dynamics, steady-state populations, relaxation time $T_1$, and decoherence time $T_2$, while numerical simulations extend the framework to concentration-dependent dynamics, $1/f$-noise-induced pure dephasing, and material-specific excitation-transfer mechanisms. Using a model system with F\"orster resonance energy transfer (FRET)-mediated excitation exchange, we show that higher qubit concentrations reduce both $T_1$ and $T_2$, whereas $1/f$ noise selectively reduces $T_2$ without changing $T_1$. Applying the framework to Er$^{3+}$-doped CeO$_2$, we find that long-range FRET-mediated excitation transfer reproduces the experimentally observed decrease in relaxation time with increasing dopant concentration, whereas short-range Dexter-type exchange does not, identifying FRET-mediated excitation transfer as the dominant interaction mechanism. The MQMF-LME framework provides a modular approach for linking microscopic interaction mechanisms and environmental noise sources to experimentally measurable decoherence times in solid-state multi-qubit systems.
\end{abstract} 

\maketitle
        
\section{Introduction}

Quantum computing has the potential to address problems that are intractable for classical computers, including molecular simulation, large-scale optimization, integer factorization, and materials modeling~\cite{qc1,qc2,qc4,qc5,qc6,qc7,qc8}. This advantage originates from the way information is stored and processed in quantum systems. Unlike classical bits, which take values of either 0 or 1, qubits can exist in superpositions of the states $|0\rangle$ and $|1\rangle$~\cite{supa}. As a result, a multi-qubit system can access a state space that grows rapidly with the number of qubits, forming the basis for quantum computational advantage. However, maintaining quantum superposition over extended times remains a major challenge. Qubits are inherently susceptible to interactions with their surrounding environment, leading to loss of coherence, or decoherence~\cite{chirolli2008decoherence,bergli2009decoherence,salamon2023qubit,gorin2007decoherence}. Understanding and mitigating decoherence is therefore a central requirement for developing practical quantum technologies. Over the past few decades, significant efforts have been devoted to exploring a wide range of quantum materials and qubit platforms, including spin qubits, semiconductor qubits, color-center qubits, quantum-dot exciton qubits, and topological qubits~\cite{russ2017three, meier2003quantum, udvarhelyi2018spin, genkin2009estimation, topo,topo2,exciton,exciton2,NVC}. These platforms differ in coherence, controllability, and scalability, but all are affected by decoherence arising from interactions with impurities and defects in the host material~\cite{studies4,studies5,bergli2009decoherence}, coupling to external fields~\cite{studies1,studies2,studies7}, strain-induced effects~\cite{studies8, studies9}, and interactions among the qubits themselves. While many studies have focused on single-qubit decoherence caused by environmental noise, the role of qubit--qubit interactions in multi-qubit systems remains less thoroughly understood.

This gap becomes particularly important in dense or highly doped qubit ensembles. A few studies of multi-qubit systems have examined quantities such as multi-qubit gate fidelity~\cite{dalton2003scaling, mqp1,mqp2,mqp3,mqp4} and entanglement measures such as concurrence~\cite{doncurrence}. However, the direct impact of qubit–qubit interactions on relaxation and decoherence remains comparatively underexplored. Experimental studies have shown that qubit performance, particularly relaxation and coherence times, can degrade with increasing qubit concentration, suggesting that inter-qubit interactions play an important role~\cite{grant2024optical,hughes2025strongly,zhang2024optical}. A predictive framework that connects qubit concentration, spatial distribution, microscopic interaction mechanism, and environmental noise to measurable relaxation and decoherence times is still lacking. Several important questions therefore remain open: How do qubits interact with one another in a dense multi-qubit environment? How strongly do these interactions contribute to the observed degradation of $T_1$ and $T_2$ with increasing qubit concentration? How does the spatial distribution of qubits modify the effective interaction rates? How do environmental noise sources combine with qubit--qubit interactions to determine the overall decoherence dynamics?

Existing theoretical and computational approaches provide important insights, but they are not always well suited for this problem. Cluster correlation expansion (CCE) methods are powerful for describing decoherence of a central spin interacting with surrounding electronic or nuclear spin baths~\cite{Giulia1, Giulia2}, but they can become computationally demanding in dense, strongly interacting environments and are often formulated around a single central spin. The Inokuti--Hirayama (IH) model has also been used to describe concentration-dependent relaxation mediated by donor--acceptor excitation transfer~\cite{inokuti1965influence}, but it was originally developed for systems with effectively unidirectional excitation transfer from excited donors to acceptors. This assumption is restrictive for multi-qubit systems composed of identical or nearly identical qubits, where excitation exchange can be bidirectional. These limitations motivate a framework that is physically interpretable, computationally tractable, and flexible enough to incorporate qubit--qubit interaction mechanisms and environmental decoherence channels.

In this work, we develop a multi-qubit mean-field Lindblad master equation (MQMF-LME) framework for modeling relaxation and decoherence in interacting multi-qubit systems. The framework is based on an open-quantum-system description, in which quantum master equations describe the time evolution of the reduced density matrix and provide a natural way to model relaxation and decoherence~\cite{zoller1997quantum,breuer2002theory,rivas2012open}. For the present goal of building a physically interpretable and computationally tractable framework, the Lindblad master equation (LME) provides a natural starting point. Quantum master equation formulations differ in their tradeoffs among accuracy, positivity, generality, and computational cost~\cite{Masterequationcompare, Masterequationcompare2}. Unlike the Redfield master equation, which can violate complete positivity~\cite{venkataraman2025coherent}, and unlike the time-convolutionless and Nakajima--Zwanzig formulations, which are designed to capture non-Markovian memory effects but can be more difficult to implement~\cite{brian2021generalized,Nonmarkov1,Nonmarkov2,shen2014exact,peng2025discretization}, the LME provides a time-local, completely positive, and computationally efficient description of Markovian open-system dynamics~\cite{gorini1976completely,lindblad1976generators,nathan2020universal}. Although the LME assumes weak system--bath coupling and short bath memory, it provides a controlled baseline for identifying how interaction-dependent transition channels contribute to relaxation and decoherence. More complex noise processes can also be incorporated through additional Lindblad operators, time-dependent rates, or stochastic Hamiltonian terms~\cite{LME_memory1,LME_memory2,LME_memory3}.

In the MQMF-LME framework, one qubit is treated as the system of interest, while the surrounding qubits are treated as an effective bath. The bath enters through interaction-dependent upward and downward transition rates, allowing the framework to describe intrinsic single-qubit relaxation, excitation transfer from the system to the bath, and excitation transfer from thermally populated bath qubits back to the system. This mean-field construction connects the dynamics to qubit concentration, spatial distribution, interaction range, and thermal bath occupation while avoiding the exponential complexity of a full many-body density-matrix treatment. A central contribution of this work is that the MQMF-LME framework provides both analytical solutions and numerical simulations. The analytical solutions yield closed-form expressions for the density-matrix dynamics, steady-state populations, relaxation time $T_1$, and decoherence time $T_2$, and establish the relaxation-limited relation $T_2 = 2T_1$ in the absence of pure dephasing. These results provide physical interpretability and serve as internal consistency checks for the numerical implementation. The simulations then extend the analytical framework to practical use cases by extracting $T_1$ and $T_2$ from density matrix dynamics, comparing different microscopic interaction mechanisms, incorporating non-exponential dephasing from \(1/f\) noise, and enabling direct comparison with experimental data on concentration-dependent relaxation dynamics.

We first apply the MQMF-LME framework to a model multi-qubit system in which qubits interact through F\"{o}rster Resonance Energy Transfer (FRET), a dipole--dipole-mediated excitation-exchange mechanism~\cite{FRET,FRET2, lakowicz2006principles}. Using this model system, we isolate the role of qubit concentration and show that increasing bath density enhances the effective transition rates, thereby reducing both $T_1$ and $T_2$. We then extend the framework to include environmental noise by introducing stochastic fluctuations in the qubit energy splitting. For Gaussian $1/f$ noise, the coherence acquires an additional non-exponential dephasing envelope, producing pure dephasing beyond the relaxation-induced decoherence caused by qubit--qubit interactions. Finally, we apply the model to Er$^{3+}$-doped CeO$_2$, a rare-earth-ion platform with an optically addressable telecom-band transition~\cite{grant2024optical, zhang2024optical, wong2024coherent}. Experimental measurements show that the relaxation time decreases with increasing Er$^{3+}$ concentration, indicating enhanced Er$^{3+}$--Er$^{3+}$ interactions~\cite{grant2024optical}. To identify the dominant interaction mechanism, we compare two limiting excitation-transfer processes: short-range Dexter-type exchange and long-range FRET-mediated dipole--dipole transfer~\cite{Dexter,FRET,FRET2,FD}. The Dexter-based model fails to reproduce the experimentally observed concentration dependence, consistent with the localized and shielded nature of the Er$^{3+}$ $4f$ orbitals. In contrast, the FRET-based MQMF-LME model reproduces the experimental relaxation trend using a physically reasonable characteristic FRET distance, identifying long-range FRET-mediated excitation transfer as the dominant mechanism responsible for concentration-dependent relaxation in Er$^{3+}$-doped CeO$_2$. Overall, the MQMF-LME framework provides a modular and generalizable approach for connecting microscopic qubit--qubit interactions and environmental noise sources to experimentally measurable relaxation and decoherence times across solid-state qubit platforms.

\section{Methods}

\begin{figure}
    \centering
    \includegraphics[width=0.2 \textwidth]{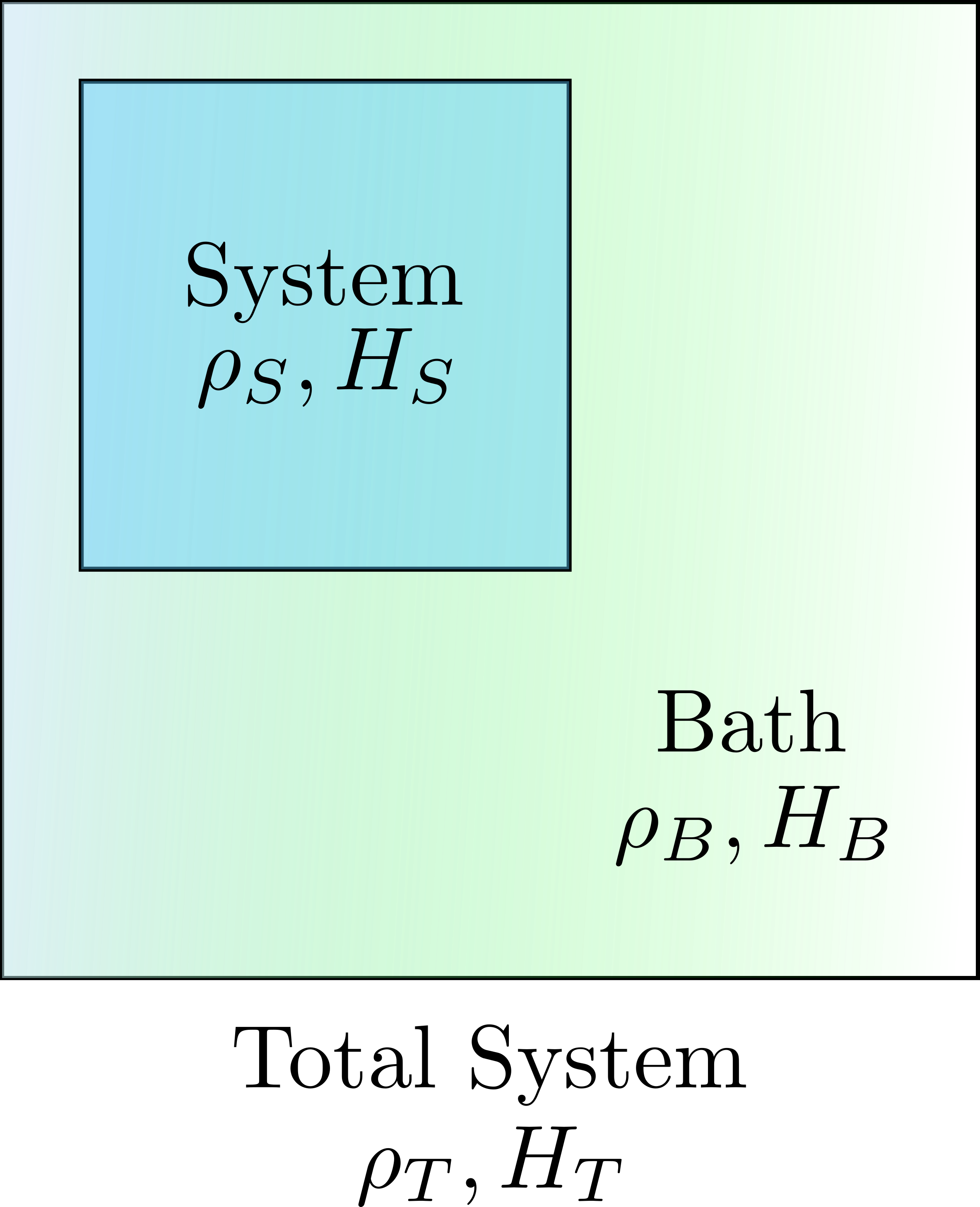} 
    \caption{{\bf Schematic of the system--bath model.} The system and the bath are described by Hamiltonians $H_S$ and $H_B$, with density matrices $\rho_S$ and $\rho_B$, respectively. The combined system is described by the total Hamiltonian $H_T$ and density matrix $\rho_T$.
}
    \label{fig:system-bath-model}
\end{figure}
 
\subsection{Lindblad Master Equation}

Here, we present a brief overview of the LME to introduce the elements relevant to our multi-qubit model; detailed derivations are available in standard references~\cite{zoller1997quantum,breuer2002theory}. We consider a system coupled to a bath, together forming the total system illustrated in Fig.~\ref{fig:system-bath-model}. The total Hamiltonian and initial density matrix are
\begin{equation}
    H_T = H_S + H_B + \alpha H_{SB} ,\quad \rho_T(0) = \rho_S(0)\otimes\rho_B,
    \label{total-hamiltonian}
\end{equation}
where $H_{S}$ and $H_{B}$ are the Hamiltonians of the system and the bath, respectively, $H_{SB}$ is the interaction Hamiltonian and $\alpha$ denotes the coupling strength.  The interaction Hamiltonian is written as 
\begin{equation}
    H_{SB} = \sum_i S_i \otimes B_i, 
    \label{eq: interaction-hamiltonian}
\end{equation} 
with $S_i$ and $B_i$ operators acting on the Hilbert spaces of the system and the bath, respectively. The evolution of the total system is governed by the von Neumann equation~\cite{Vonneumann}:
\begin{equation}
\dot{\rho}_T(t) = -i[H_T, \rho_T], 
\label{eq:VonNeuman}
\end{equation}
which describes unitary evolution of the total system under its Hamiltonian $H_T$.

We are primarily interested in analyzing the time evolution of the system alone. Performing a partial trace of Eq.~\ref{eq:VonNeuman} over the bath degrees of freedom yields a master equation for the reduced density matrix, $\rho_S(t)$. The most general completely positive and trace-preserving form of such Markovian evolution is the LME. The LME provides a consistent framework for modeling dissipation in open quantum systems: it guarantees complete positivity and trace preservation, represents physical processes through appropriate Lindblad operators, and remains linear and first order in the density matrix--properties that enable analytical solutions in simple cases and efficient numerical treatment in more complex scenarios. The LME is written as
\begin{equation}
\begin{split}
    \frac{d \rho_S(t)}{dt} 
    &= -i[H_S + H_{LS}, \rho_S(t)] \\
    &\hspace{-1.0 cm}+\sum_{i,\omega} \Big(L_i(\omega)\rho_S(t) L_i^{\dagger}(\omega) 
      - \tfrac{1}{2}\{ L_i^{\dagger}(\omega) L_i(\omega), \rho_S(t)\} \Big), 
\end{split}
\label{eq:LME}
\end{equation}
where $\omega$ defines the energy difference between eigenstates of $H_S$ and the sum over \(\omega\) accounts for all system transitions induced by the bath. The Lindblad operators are $L_i (\omega) = \sqrt{\Gamma_i (\omega)} S_i (\omega)$ with $S_i (\omega)$ characterizing the system--bath coupling associated with a transition of energy \(\omega\), and $\Gamma_i (\omega)$ denoting the corresponding rates. The first term accounts for coherent evolution generated by the system Hamiltonian, including the Lamb-shift correction, $H_{LS}$, while the second term describes dissipative processes induced by environmental interactions. A complete derivation of the LME is provided in Appendix~\ref{app:LME}.

\subsection{Interacting Qubit Model: Two-Qubit System}

\begin{figure}
    \centering
    \includegraphics[width=0.35 \textwidth]{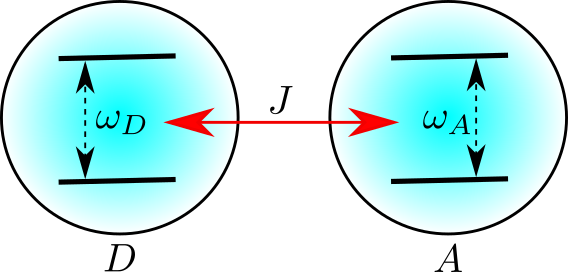} 
    \caption{{\bf Two-qubit interaction model.} Schematic of a two-qubit system, where each qubit is treated as a two-level system with transition energies $\omega_D$ and $\omega_A$. The coherent exchange pathway can be represented by a coupling strength \(J\), however, the dissipative formulation used here represents excitation transfer through Lindblad jump operators.
    }
    \label{fig:two-qubits}
\end{figure}

As a first step toward modeling a multi-qubit environment, we consider an interacting two-qubit system, as shown in Fig.~\ref{fig:two-qubits}. We model the qubit--qubit interaction as a FRET-mediated excitation-transfer process arising from non-radiative near-field dipole--dipole coupling. Such excitation transfer can, in principle, be represented coherently through an exchange Hamiltonian of the form
\begin{equation}
H = \frac{\omega_D}{2}\sigma_z^D + \frac{\omega_A}{2}\sigma_z^A + J(\sigma_+^D\sigma_-^A + \sigma_+^A\sigma_-^D),
\label{eq:hamiltonian-two-qubit}
\end{equation}
where \(D\) and \(A\) label the two interacting qubits and may be interpreted as donor-like and acceptor-like sites in a FRET-type excitation-transfer process. The first two terms represent the corresponding energy splittings, with $\omega_D$ and $\omega_A$ denoting the transition energies of qubits \(D\) and \(A\), respectively. \(\sigma_z^D\) and \(\sigma_z^A\) are Pauli $z$ operators acting on qubits \(D\) and \(A\), respectively. The ladder operators \(\sigma_+^D\), \(\sigma_-^D\), \(\sigma_+^A\), and \(\sigma_-^A\) create and annihilate excitations on the corresponding qubits. The final term describes coherent excitation exchange between the qubits, with $J$ quantifying the coherent exchange strength. The operator $\sigma_+^D\sigma_-^A$ corresponds to transfer of an excitation from qubit \(A\) to qubit \(D\), while $\sigma_+^A\sigma_-^D$ represents the reverse process (from \(D\) to \(A\)).

In the dissipative formulation used in this work, however, the excitation-transfer process is represented by Lindblad jump operators rather than by the coherent exchange coupling \(J\). We therefore introduce Eq.~\ref{eq:hamiltonian-two-qubit} only to distinguish coherent exchange from the incoherent transfer processes used below. To avoid double counting the same transfer mechanism, the coherent part of the LME is taken to be the free two-qubit Hamiltonian
\begin{equation}
H_S = \frac{\omega_D}{2}\,\sigma_z^{D} + \frac{\omega_A}{2}\,\sigma_z^{A}.
\label{eq:free-two-qubit-Hamiltonian}
\end{equation}
To model dissipative excitation transfer between the two qubits, we use the LME
\begin{equation}
\frac{d\rho}{dt} = -i[H_S,\rho] + \mathcal{D}[L_{AD}]\rho + \mathcal{D}[L_{DA}]\rho,
\label{eq:LME-two-qubit}
\end{equation}
where \(\rho\) is the density matrix of the two-qubit system and $H_S$ is the free two-qubit Hamiltonian defined in Eq.~\ref{eq:free-two-qubit-Hamiltonian}. The Lindblad dissipator is defined as $\mathcal{D}[L]\rho=L \rho L^{\dagger} - \tfrac{1}{2}\{ L^{\dagger}L, \rho\}$. The operators $L_{AD} = \sqrt{\gamma_{AD}}\sigma_+^D\sigma_-^A$ and $L_{DA} = \sqrt{\gamma_{DA}}\sigma_+^A\sigma_-^D$ describe incoherent excitation transfer from qubit \(A\) $\rightarrow$ \(D\) and \(D\) $\rightarrow$ \(A\), respectively, with $\gamma_{AD}$ and $\gamma_{DA}$ denoting the corresponding transfer rates. The dissipators therefore take the explicit forms
\begin{equation}
    \hspace{-0.3 cm}\mathcal{D}[L_{AD}]\rho = \gamma_{AD}\left(\sigma_+^D\sigma_-^A\rho\sigma_+^A\sigma_-^D - \frac{1}{2}\left\{\sigma_+^A\sigma_-^D\sigma_+^D\sigma_-^A, \rho\right\}\right)
    \label{eq:dissipators-two-qubits_2}
\end{equation}
\begin{equation}
\hspace{-0.3 cm}\mathcal{D}[L_{DA}]\rho = 
\gamma_{DA}\left(\sigma_+^A\sigma_-^D\rho\sigma_+^D\sigma_-^A - \frac{1} {2}\left\{\sigma_+^D\sigma_-^A\sigma_+^A\sigma_-^D, \rho\right\}\right)
\label{eq:dissipators-two-qubits_1}
\end{equation}
Assuming the qubits interact with a thermal environment, the transition rates satisfy detailed balance,
\begin{equation}
\frac{\gamma_{DA}}{\gamma_{AD}} = e^{(\omega_D - \omega_A)/k_BT}.
\label{eq:thermal-balance-two-qubits}
\end{equation}
For degenerate qubits ($\omega_D = \omega_A$), the transitions become symmetric and the rates satisfy $\gamma_{AD} = \gamma_{DA}$.

\subsection{Interacting Qubit Model: Multi-Qubit System}

\begin{figure}
    \centering
    \includegraphics[width=0.35 \textwidth]{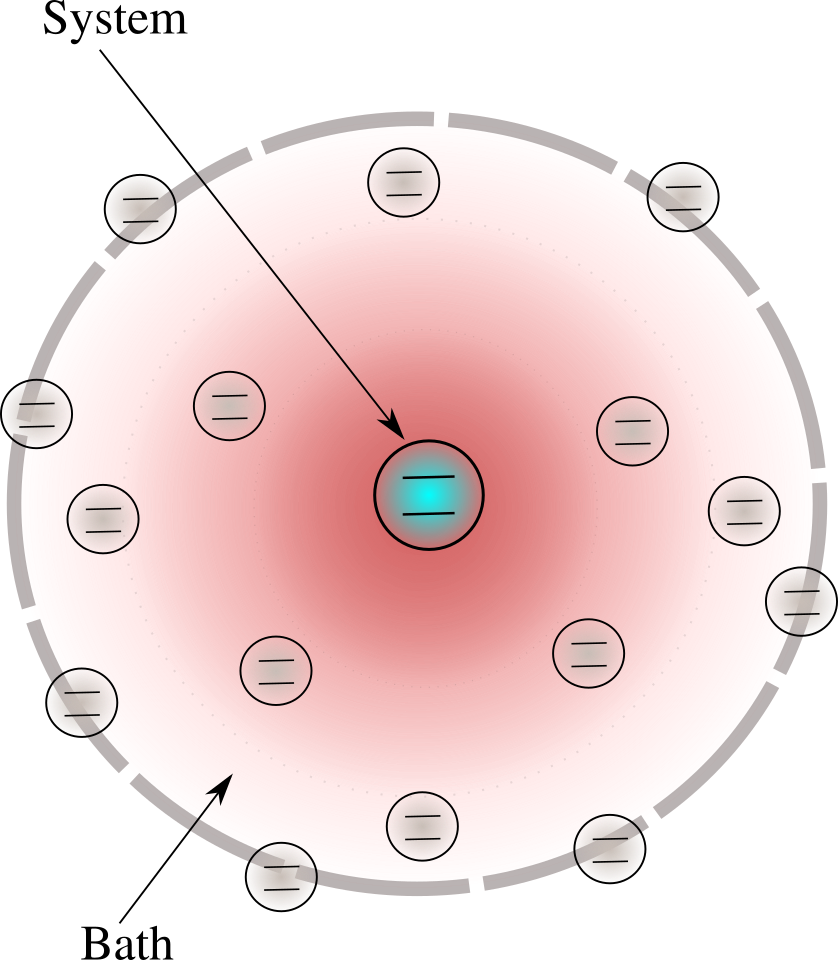} 
    \caption{{\bf Multi-qubit interaction model.} Schematic of a multi-qubit system in which one qubit (blue) is designated as the system and the surrounding qubits form an effective bath (gray region). The system qubit is characterized by an energy splitting $\omega_0$.
}
    \label{fig:bath}
\end{figure}

Two-qubit models provide a useful starting point for understanding decoherence arising from qubit--qubit interactions, but real systems often contain many qubits within a finite volume. In such dense environments, the dynamics become more complex because each qubit experiences the combined influence of many neighboring qubits, and keeping track of all pairwise couplings quickly becomes intractable. To capture these collective effects without resorting to full many-body simulations, we adopt a mean-field description of an ensemble of $N+1$ interacting qubits within the LME framework. We refer to this approach as the multi-qubit mean-field Lindblad master equation (MQMF-LME) framework. In this approach, we designate one qubit as the system of interest and treat the remaining $N$ qubits as an effective bath, as illustrated in Fig.~\ref{fig:bath}. This coarse-grained description preserves the essential interaction physics while avoiding the exponential complexity of the full problem. Within the MQMF-LME framework, the system experiences two primary decoherence mechanisms: (1) intrinsic relaxation of the system qubit and (2) excitation exchange between the system qubit and the surrounding bath qubits.

We model the bath-assisted excitation-exchange channel using FRET-mediated dipole--dipole interactions between the system qubit and the bath qubits. Because FRET is a near-field dipole--dipole process, its strength decays rapidly with separation, following the characteristic $R^{-6}$ dependence~\cite{forster1965}. Accordingly, the coupling between the system qubit and the $i^{\text{th}}$ bath qubit is characterized by the rate
\begin{equation}
\gamma_{0i} = \gamma_0 \left(\frac{R_0}{R_{0i}}\right)^6, 
\label{eq:gamma-FRET}
\end{equation}
where $R_{0i}$ is the distance between the system qubit (labeled 0) and $i^{\text{th}}$ bath qubit, $R_0$ is the characteristic FRET distance, and $\gamma_0$ sets the overall interaction scale. We assume that the bath is in thermal equilibrium at temperature $T$. Under this assumption, excitation transfer is bidirectional: the system can transfer excitation to the bath, and thermally populated bath states can transfer excitation back to the system. This bidirectional FRET-mediated excitation exchange defines the bath-induced transition channels, \(\Gamma_\uparrow, \Gamma_\downarrow\), used in the MQMF-LME framework.

The bath-induced downward and upward transition rates of the system qubit are obtained by summing over the contributions from all bath qubits:
\begin{align}
&\Gamma_{\downarrow} = \sum_{i=1}^N \gamma_{0i} (1 - P_i^{e}) \quad \text{and} \\
&\Gamma_{\uparrow} = \sum_{i=1}^N \gamma_{i0} P_i^{e},
\label{eq:Lindblad Dissipator}
\end{align}
where $P_i^{e} = [1 + \exp(\omega_i/k_BT)]^{-1}$ is the excited-state probability of the $i^{\text{th}}$ bath qubit~\cite{fermifundamentals}, $\omega_i$ is its energy splitting, $k_B$ is the Boltzmann constant and $T$ is the bath temperature. Here, $\Gamma_{\downarrow}$ describes excitation loss from the system to the bath, whereas $\Gamma_{\uparrow}$ describes excitation gain from the bath to the system. For identical qubits ($\omega_i = \omega_0$) and symmetric coupling ($\gamma_{i0} = \gamma_{0i}$); therefore, the upward transition rate reduces to
\begin{equation}
\Gamma_{\uparrow} = \sum_{i=1}^N \gamma_{0i} \frac{1}{1 + e^{\omega_0/k_BT}}.
\label{eq:Dissipator with fermionic statistics}
\end{equation}
To evaluate these rates, we replace the discrete sum with a continuum average over the spatial distribution of bath qubits. Assuming that the bath qubits are randomly and uniformly distributed in a volume $V$, the number of bath qubits in a spherical shell between \(R\) and \(R+dR\) is $C 4\pi R^2 dR$, where \(C=N/V\) is the qubit concentration. The discrete sum over bath qubits can therefore be converted into an integral over inter-qubit distance, yielding the thermally averaged transition rates
\begin{equation}
\Gamma_{\uparrow} = C\gamma_0 R_0^6 \left(\frac{1}{1+e^{\omega_0/k_BT}}\right) \int_{R_{min}}^{R_{max}} \frac{4\pi R^2}{R^6} dR
\label{eq:Abosorption dissipator with meanfiled framework}
\end{equation}
\begin{equation}
\Gamma_{\downarrow} = C\gamma_0 R_0^6 \left(\frac{e^{\omega_0/k_BT}}{1+e^{\omega_0/k_BT}}\right) \int_{R_{min}}^{R_{max}} \frac{4\pi R^2}{R^6} dR
\label{eq:Relaxation dissipator with meanfiled framework}
\end{equation}
where $R_{min}$ is the minimum allowed separation between qubits and $R_{max}$ is the FRET cutoff distance. The thermal factors satisfy detailed balance: 
\begin{equation}
    \frac{\Gamma_{\downarrow}}{\Gamma_{\uparrow}} = e^{\omega_0 / k_BT}
    \label{eq: detailed balance}
\end{equation}
The distance integral can be evaluated analytically, 
\begin{equation}
\int_{R_{\min}}^{R_{\max}} \frac{4\pi R^2}{R^6}dR = \frac{4\pi}{3} \left(\frac{1}{R_{\min}^3} -\frac{1}{R_{\max}^3} \right).
\end{equation}
Defining
\begin{equation}
K_0 = C\gamma_0R_0^6 \frac{4\pi}{3} \left( \frac{1}{R_{\min}^3}-\frac{1}{R_{\max}^3} \right),
\end{equation}
the upward and downward transition rates become
\begin{equation}
\Gamma_{\uparrow} = \frac{K_0}{1+e^{\omega_0/k_BT}}, \qquad
\Gamma_{\downarrow} = \frac{K_0e^{\omega_0/k_BT}}{1+e^{\omega_0/k_BT}}.
\label{eq:closed-form-rates}
\end{equation}
Therefore, the total FRET-induced transition rate is
\begin{equation}
\Gamma_{\uparrow}+\Gamma_{\downarrow} = K_0 = C\gamma_0R_0^6 \frac{4\pi}{3} \left(
\frac{1}{R_{\min}^3} - \frac{1}{R_{\max}^3} \right).
\label{eq:total-FRET-rate}
\end{equation}
Thus, for fixed \(C, R_0,R_{\min}, R_{\max},\) and \(\gamma_0\), temperature redistributes the relative strengths of the upward and downward transition rates but does not change their sum.

\begin{figure}
    \centering
    \includegraphics[width=0.45\textwidth]{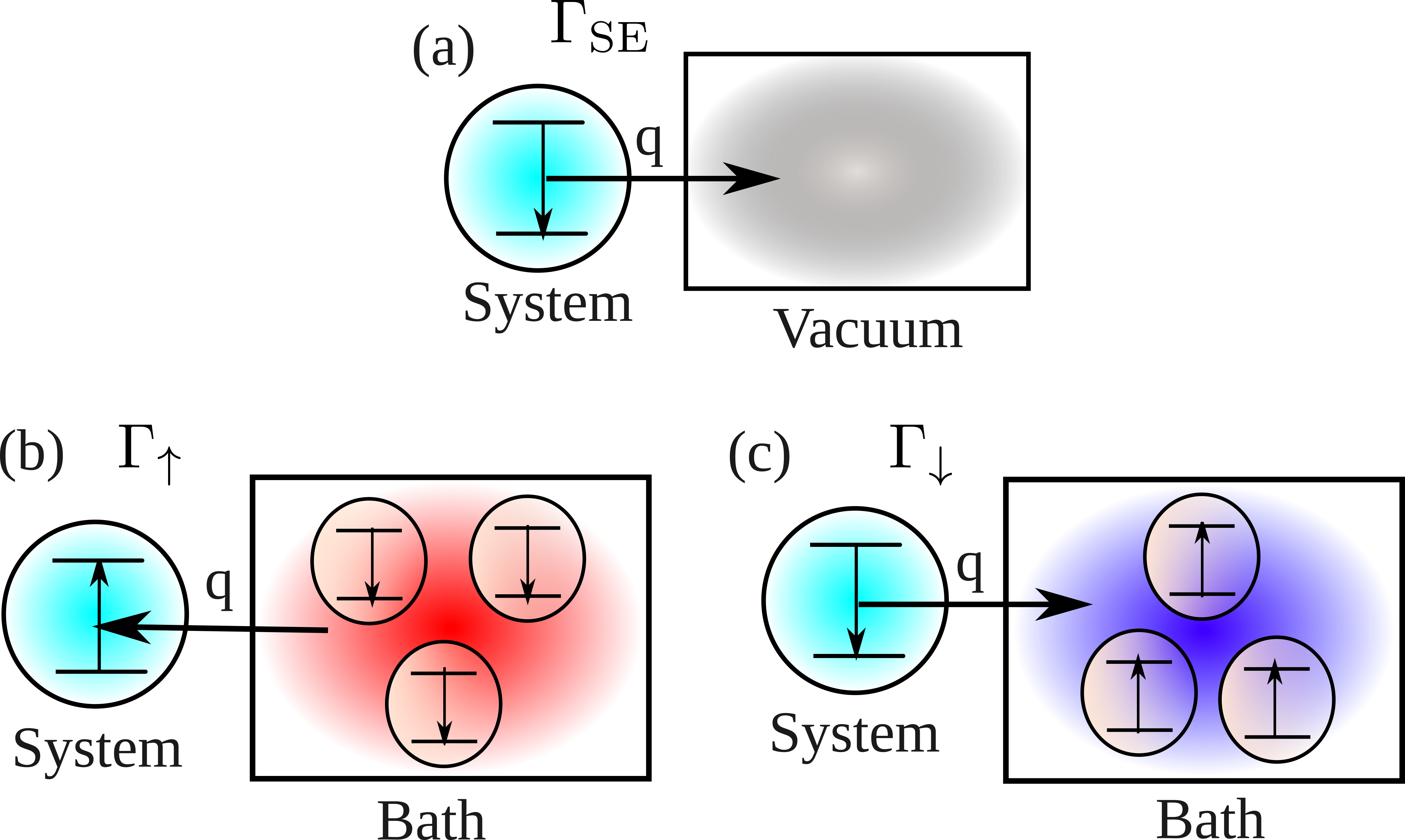} 
    \caption{\textbf{Dissipation channels in the multi-qubit model.} (a) Intrinsic single-qubit relaxation with rate $\Gamma_{\text{SE}}$. (b) Excitation transfer from the bath to the system with rate $\Gamma_{\uparrow}$. (c) Excitation transfer from the system to the bath with rate $\Gamma_{\downarrow}$.}
    \label{fig:int}
\end{figure}

We initialize the system qubit in a superposition state \( |\psi\rangle \) defined as:
\begin{equation}
    |\psi\rangle = \frac{1}{\sqrt{2}}\left( |e\rangle + |g\rangle \right).
    \label{eq:Superposition state}
\end{equation}
The corresponding density matrix is
\begin{equation}
    \rho = 
    \begin{bmatrix}
        \rho_{ee} & \rho_{eg} \\
        \rho_{ge} & \rho_{gg}
    \end{bmatrix}.
    \label{eq:Density matrix of superposition state}
\end{equation}
The time evolution of the density matrix is governed by the LME
\begin{align}
\frac{d \rho}{d t} =\;& -i [H, \rho] \nonumber \\
&+ \Gamma_{\mathrm{\text{SE}}} \left( \sigma_- \rho \sigma_+ - \frac{1}{2} \left\{ \sigma_+\sigma_-, \rho \right\} \right) \nonumber \\
&+ \Gamma_{\downarrow} \left( \sigma_- \rho \sigma_+ - \frac{1}{2} \left\{ \sigma_+ \sigma_-, \rho \right\} \right) \nonumber \\
&+ \Gamma_{\uparrow} \left( \sigma_+ \rho \sigma_- - \frac{1}{2} \left\{ \sigma_- \sigma_+, \rho \right\} \right).
\label{eq:Complete-Lindblad-Master-Equation-for-multiqubit-system}
\end{align}
Here, \(H = \frac{\omega_0}{2} \sigma_z\) is the system Hamiltonian, \(\omega_0\) is the energy splitting, \(\sigma_z\) is the Pauli \(z\) operator, and $\Gamma_{\text{SE}}$ represents the intrinsic spontaneous-emission rate. Figure~\ref{fig:int} summarizes the dissipation channels included in the MQMF-LME framework: intrinsic relaxation, excitation transfer from the bath to the system, and excitation transfer from the system to the bath. The decay of the off-diagonal elements \(\rho_{eg}\) and \(\rho_{ge}\) characterizes the loss of coherence of the superposition state, while the decay of the diagonal element \(\rho_{ee}\) describes relaxation of the excited-state population.
For compact notation, we define
\begin{equation}
\Gamma_-=\Gamma_{\mathrm{SE}}+\Gamma_{\downarrow},
\qquad
\Gamma_+=\Gamma_{\uparrow}.
\end{equation}
With these definitions, the MQMF-LME yields the following coupled equations for the density-matrix elements:
\begin{equation}
\dot\rho_{ee} = -\Gamma_-\rho_{ee}+\Gamma_+\rho_{gg},
\label{eq:Differential-equation-for-excited-state-population}
\end{equation}
\begin{equation}
\dot\rho_{gg} = \Gamma_-\rho_{ee}-\Gamma_+\rho_{gg},
\label{eq:Differential-equation-for-ground-state-population}
\end{equation}
\begin{equation}
\dot\rho_{eg} = -i\omega_0\rho_{eg}-\frac{\Gamma_+ + \Gamma_-}{2}\rho_{eg},
\label{eq:Differential-equation-for-off-diagonal-elements-2}
\end{equation}
and
\begin{equation}
\dot\rho_{ge} = i\omega_0\rho_{ge}-\frac{\Gamma_+ + \Gamma_-}{2}\rho_{ge}.
\label{eq:Differential-equation-for-off-diagonal-elements-1}
\end{equation}
These equations show explicitly that the population relaxes toward its steady-state value with rate \((\Gamma_+ + \Gamma_-)\), while the coherence amplitude decays at half this rate in the absence of pure dephasing. Solving the first-order differential equations gives 
\begin{equation*}
    \rho_{ee}(t) = \frac{\Gamma_+}{\Gamma_+ + \Gamma_-}+\left( \rho_{ee}(0)-\frac{\Gamma_+}{\Gamma_+ + \Gamma_-}\right)e^{-(\Gamma_+ +\Gamma_-)t},
    \label{eq:Analytical-Solution-for-excited-state-population}
\end{equation*}
\begin{equation*}
    \rho_{gg}(t) = \frac{\Gamma_-}{\Gamma_+ + \Gamma_-}+\left( \rho_{gg}(0)-\frac{\Gamma_-}{\Gamma_+ + \Gamma_-}\right)e^{-(\Gamma_+ +\Gamma_-)t},
    \label{eq:Analytical-Solution-for-ground-state-population}
\end{equation*}
\begin{equation*}
    \rho_{ge}(t) = \rho_{ge}(0) e^{i\omega_0t}e^{-\left({\frac{\Gamma_+ +\Gamma_-}{2}}\right)t},
    \label{eq:Analytical-Solution-for-off-diagonal-element-1}
\end{equation*}
and
\begin{equation}
    \rho_{eg}(t) = \rho_{eg}(0) e^{-i\omega_0t}e^{-\left({\frac{\Gamma_+ +\Gamma_-}{2}}\right)t}.
    \label{eq:Analytical-solution-for-off-diagonal-element-2}
\end{equation}
A detailed derivation of these analytical solutions is provided in Appendix~\ref{app:analytical}.

The steady-state populations follow from the long-time limit of the diagonal solutions:
\begin{equation}
\rho_{ee}^{\mathrm{ss}} = \frac{\Gamma_+}{\Gamma_+ + \Gamma_-} = \frac{\Gamma_{\uparrow}}{\Gamma_{\mathrm{SE}}+\Gamma_{\downarrow}+\Gamma_{\uparrow}},
\end{equation}
and
\begin{equation}
\rho_{gg}^{\mathrm{ss}}= \frac{\Gamma_-}{\Gamma_+ + \Gamma_-} = \frac{\Gamma_{\mathrm{SE}} +\Gamma_{\downarrow}}{\Gamma_{\mathrm{SE}}+\Gamma_{\downarrow}+\Gamma_{\uparrow}}.
\end{equation}
Thus, when upward excitation transfer from the bath is present, the excited-state population relaxes toward a finite steady-state value rather than decaying completely to zero.

The same analytical solutions define the relaxation and decoherence times used in the Results section. The population relaxation rate is
\begin{equation}
\frac{1}{T_1} = \Gamma_+ + \Gamma_- = \Gamma_{\mathrm{SE}}+\Gamma_{\downarrow}+\Gamma_{\uparrow}.
\end{equation}
In the absence of an explicit pure-dephasing channel, the transverse decoherence rate is
\begin{equation}
\frac{1}{T_2} = \frac{1}{2T_1} = \frac{\Gamma_+ + \Gamma_-}{2},
\end{equation}
or equivalently \(T_2 =2 T_1\). Therefore, deviations from \(T_2 = 2T_1 \) require an additional dephasing mechanism or a temperature-dependent total interaction rate. If pure dephasing is included as an additional channel, this relation generalizes to~\cite{burkard2004multilevel,falci2005initial}
\begin{equation}
\frac{1}{T_2} = \frac{1}{2T_1} + \frac{1}{T_\phi},
\end{equation}
where \(T_\phi\) is the pure dephasing time. 

\subsection{Noise-Induced Dephasing in Multi-Qubit Systems}

In addition to decoherence arising from qubit--qubit interactions, multi-qubit systems are subject to environmental fluctuations that introduce additional dephasing channels. In solid-state qubit platforms, common noise sources include defect states in the host material~\cite{qubitdefect}, vacancy centers, trapped charges~\cite{wilen2021correlated, christensen2019anomalous,kuhlmann2013charge}, and fluctuating external electric and magnetic fields~\cite{wei2022measurement}. These fluctuations can induce temporal variations in the qubit energy splitting, $\omega_0$, leading to stochastic modulation of the qubit transition energy~\cite{dephasing4, dephasing6}. Such stochastic fluctuations give rise to pure dephasing, reducing quantum coherence without directly affecting the population dynamics. To account for this mechanism, we extend the MQMF-LME framework by introducing a stochastic fluctuation in the qubit energy splitting. The resulting time-dependent Hamiltonian is written as: 
\begin{equation}
H(t)= \frac{\omega_0+f(t)}{2}\sigma_z.
\end{equation}
Here, $\omega_0$ is the unperturbed qubit energy splitting and $f(t)$ describes the stochastic fluctuations induced by the surrounding noisy environment. The specific form of $f(t)$ depends on the underlying noise model and may represent Gaussian noise, random-telegraph noise, or $1/f$ noise. Because the fluctuation couples through the $\sigma_z$ operator, it modulates the qubit energy splitting without directly driving transitions between the ground and excited states. Consequently, the primary effect of $f(t)$ is pure dephasing. This dephasing channel contributes to \(T_2\) but does not affect \(T_1\). Including this stochastic Hamiltonian, the extended MQMF-LME can be expressed as
\begin{align}
\frac{d \rho}{d t} =\;& -i [H(t), \rho] \nonumber \\
&+ \Gamma_{\mathrm{\text{SE}}} \left( \sigma_- \rho \sigma_+ - \frac{1}{2} \left\{ \sigma_+\sigma_-, \rho \right\} \right) \nonumber \\
&+ \Gamma_{\downarrow} \left( \sigma_- \rho \sigma_+ - \frac{1}{2} \left\{ \sigma_+ \sigma_-, \rho \right\} \right) \nonumber \\
&+ \Gamma_{\uparrow} \left( \sigma_+ \rho \sigma_- - \frac{1}{2} \left\{ \sigma_- \sigma_+, \rho \right\} \right).
\label{eq:Complete-Lindblad-Master-Equation-for-multiqubit-system-for-dephasing}
\end{align}
The dissipative terms remain unchanged from the original MQMF-LME (Eq.~\ref{eq:Complete-Lindblad-Master-Equation-for-multiqubit-system}) and continue to describe intrinsic single-qubit relaxation, excitation transfer from the system to the bath, and excitation transfer from the bath to the system. The environmental noise enters through the stochastic Hamiltonian contribution $f(t)\sigma_z/2$. As a result, the phase accumulated by the qubit depends on the particular noise realization. For an individual realization of \(f(t)\), this contribution produces phase modulation. After ensemble averaging over many noise realizations, the accumulated random phase leads to decay of the off-diagonal density-matrix elements, thereby producing pure dephasing.

We next derive the ensemble-averaged coherences of the extended MQMF-LME by considering \(1/f\) noise as a representative source of pure dephasing. Because the noise term couples through \(\sigma_z\), it modulates the energy splitting of the qubit without inducing transitions between the ground and excited states. As a result, the diagonal density matrix elements remain identical to those obtained from the original MQMF-LME. In contrast, the off-diagonal elements acquire an additional dephasing factor, $e^{-\chi(t)}$, which accounts for the loss of coherence arising from stochastic fluctuations in the qubit energy splitting. The resulting ensemble-averaged coherences are given by
\begin{eqnarray}
\langle \rho_{eg}(t)\rangle=
\rho_{eg}(0)
e^{-i\omega_0 t}
e^{-\frac{\Gamma_-+\Gamma_+}{2}t}
e^{-\chi(t)}, \nonumber \\
\langle \rho_{ge}(t)\rangle
=
\rho_{ge}(0)
e^{i\omega_0 t}
e^{-\frac{\Gamma_-+\Gamma_+}{2}t}
e^{-\chi(t)}.
\label{eq:ensemble-averaged-analytical-solution-for-density-matrix-elements}
\end{eqnarray}
The first exponential factor describes the coherent phase evolution at the qubit energy splitting $\omega_0$. The second factor represents relaxation-induced coherence decay 
arising from the dissipative channels in the MQMF-LME framework. The final factor, $e^{-\chi(t)}$, captures the additional loss of phase coherence due to noise-induced dephasing. For a single realization of the stochastic fluctuation \(f(t)\), the accumulated random phase is
\begin{equation}
\phi(t)=\int_0^t f(t')\,dt'.
\label{eq:stochastic-phase}
\end{equation}
After averaging over many noise realizations, the random phase factor becomes
\begin{equation}
\left\langle e^{\pm i\phi(t)}\right\rangle = e^{-\chi(t)},
\label{eq:dephasing-factor}
\end{equation}
where \(\chi(t)\) is the dephasing function. For zero-mean Gaussian noise, this function is determined by the noise autocorrelation,
\begin{equation}
\chi(t) = \frac{1}{2} \int_0^t dt_1 \int_0^t dt_2 \left\langle f(t_1)f(t_2)\right\rangle.
\label{eq:chi-autocorrelation}
\end{equation}
The autocorrelation function describes how strongly fluctuations at different times remain correlated and therefore determines the rate at which phase coherence is lost. Environmental noise, however, is often characterized experimentally in the frequency domain through its power spectral density. By invoking the Wiener--Khinchin theorem~\cite{DPS}, the dephasing function can be expressed in terms of the power spectral density $S_f(\omega)$ as,
\begin{equation}
\chi(t) = \frac{1}{\pi} \int_0^\infty d\omega\, S_f(\omega) \frac{1-\cos(\omega t)}{\omega^2}.
\label{eq:filter_function_integral}
\end{equation}
Here $S_f(\omega)$ is the power spectral density of the stochastic fluctuation \(f(t)\). For \(1/f\) noise, the power spectral density is assumed to follow
\[
S_f(\omega)=\frac{A}{|\omega|},
\]
where $A$ characterizes the noise strength. Because an ideal $1/f$ spectrum diverges in both the low-frequency and high-frequency limits, it is necessary to introduce physically motivated cutoffs. We denote the low- and high-frequency cutoffs by $\omega_l$ and $\omega_h$, respectively. These cutoffs account for the finite observation time and the finite bandwidth of the microscopic noise sources, ensuring that the noise spectrum remains well defined. The corresponding \(1/f\)-noise dephasing function is
\begin{equation}
\chi_{1/f}(t) = \frac{A}{\pi} \int_{\omega_l}^{\omega_h} d\omega\, \frac{1-\cos(\omega t)}{\omega^3}.
\end{equation}
In the experimentally relevant regime $\omega_l t \ll 1$ and $\omega_h t \gg 1$, the dominant contribution to dephasing comes from the low-frequency part of the noise spectrum. Under these conditions, the dephasing function can be approximated as
\begin{equation}
\chi_{1/f}(t) \approx \frac{A t^2}{2\pi} \ln\left(\frac{1}{\omega_l t}\right).
\label{eq:chi-one-over-f-asymptotic}
\end{equation}
Thus, \(1/f\) noise introduces a non-exponential dephasing envelope, \(e^{-\chi_{1/f}(t)}\), in addition to the exponential coherence decay caused by the Lindblad relaxation channels. If the noise-induced coherence decay is approximately exponential, the effect of noise can be summarized by a pure-dephasing time \(T_\phi \). More generally, for \(1/f\) noise, the coherence decay is described by the full dephasing factor \(e^{-\chi_{1/f}(t)}\). A detailed derivation of these analytical expressions is provided in Appendix~\ref{app:analytical}.

\section{Results}

\begin{figure}
    \centering
    \includegraphics[width=0.4\textwidth]{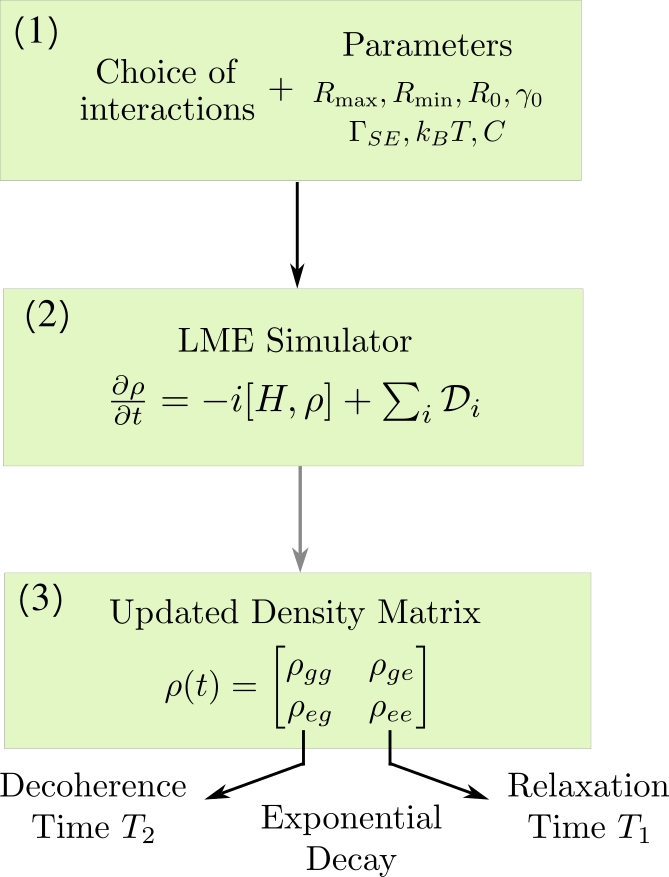} 
    \caption{\textbf{MQMF-LME simulation workflow}. (1) Input parameters defining the multi-qubit system are specified, including the interaction mechanism, qubit concentration, bath temperature, interaction range, and characteristic interaction scale. (2) These inputs are used to construct the transition rates, $\Gamma_{\uparrow}$ and $\Gamma_{\downarrow}$, that define the dissipative dynamics in the MQMF-LME framework. (3) The master equation is solved to obtain the time evolution of the density matrix. \(T_1\) and \(T_2\) are extracted from the relaxation of the excited-state population \(\rho_{ee}(t)\) and the decay of the coherence \(\rho_{eg}(t)\), respectively.
}
    \label{fig:flow}
\end{figure}

In this section, we present the relaxation and decoherence dynamics predicted by the MQMF-LME framework for both a model multi-qubit system and a realistic rare-earth-ion multi-qubit platform. We first apply the framework to a simplified model system to isolate the effect of qubit concentration on system relaxation and decoherence dynamics. We then introduce \(1/f\) noise to examine how an additional pure-dephasing channel modifies the coherence dynamics without changing the population relaxation. Finally, we extend the analysis to Er$^{3+}$-doped CeO$_2$, where concentration-dependent relaxation has been experimentally observed ~\cite{grant2024optical}. This latter case study allows us to evaluate whether the framework reproduces the measured relaxation trends and to identify the dominant inter-qubit excitation-transfer mechanism responsible for the experimentally observed relaxation behavior. Figure~\ref{fig:flow} summarizes the implementation of the MQMF-LME framework and the associated simulation workflow. First, for a given multi-qubit system, we identify the relevant qubit–qubit interaction mechanism and specify the model parameters: the interaction range ($R_{\min}$ and $R_{\max}$), the characteristic interaction distance $R_0$, the bath temperature $T$, the qubit concentration $C$, and the spatial distribution of qubits. The parameter $R_0$ sets the interaction scale and depends on the microscopic transfer mechanism and the host-material environment, including spectral overlap, orbital overlap, transition dipole strength, dielectric screening, and the characteristic interaction timescale. Here, we consider both long-range and short-range interactions: FRET represents long-range dipole--dipole-mediated excitation transfer~\cite{lakowicz2006principles,FRET2}, whereas Dexter-type transfer represents short-range exchange-mediated excitation transfer~\cite{skourtis2016dexter, clegg1995fluorescence, FRET2, skourtis2016dexter}. Second, using these inputs, we construct the corresponding interaction-dependent transition rates ($\Gamma_\uparrow$, $\Gamma_\downarrow$) that govern the dissipative dynamics within the MQMF-LME framework. Finally, we solve the resulting master equation to obtain the time evolution of the density matrix, whose elements directly encode the relevant timescales of the qubit dynamics. We extract the relaxation time $T_1$ from the excited-state population $\rho_{ee}(t)$ and the decoherence time $T_2$ from the coherence $\rho_{eg}(t)$.

\subsection{Model Multi-Qubit System}

We first apply the MQMF-LME framework to a  simplified model system composed of qubits with an energy splitting of $\omega_0 =0.07$ eV between the ground and excited states and consider FRET-mediated excitation exchange between neighboring qubits. Because FRET is typically effective over donor--acceptor separations of approximately \(1\)–\(10\,\text{nm}\) ~\cite{sahoo2011forster}, we introduce lower and upper cutoff distances, \(R_{\min}=0.5\,\text{nm}\) and \(R_{\max}=5\,\text{nm}\), respectively, to define the range of inter-qubit interactions included in the simulation. The lower cutoff is chosen to be comparable to the sum of the effective atomic radii of donor and acceptor atoms, while the upper cutoff is selected to remain within the physically relevant regime for FRET interactions~\cite{andrews1999resonance}. We choose a characteristic F\"{o}rster radius of \(R_0 = 1.5\,\mathrm{nm}\), consistent with typical values reported for FRET systems ~\cite{andrews1999resonance}. The parameter \( \gamma_0 \) sets the characteristic timescale of the FRET-mediated dynamics, while \(\Gamma_{\text{SE}}\) denotes the intrinsic single-qubit relaxation rate. In this work, we set \( \gamma_0 = \Gamma_{\text{SE}} = 100\,\mathrm{s}^{-1} \),  which corresponds to relaxation processes on the millisecond time scale~\cite{grant2024optical}. This choice provides a physically reasonable timescale for exploring interaction-driven relaxation and decoherence of a single qubit in a multi-qubit system within the MQMF-LME framework. 

\begin{figure}
    \centering
    \includegraphics[width=0.4\textwidth]{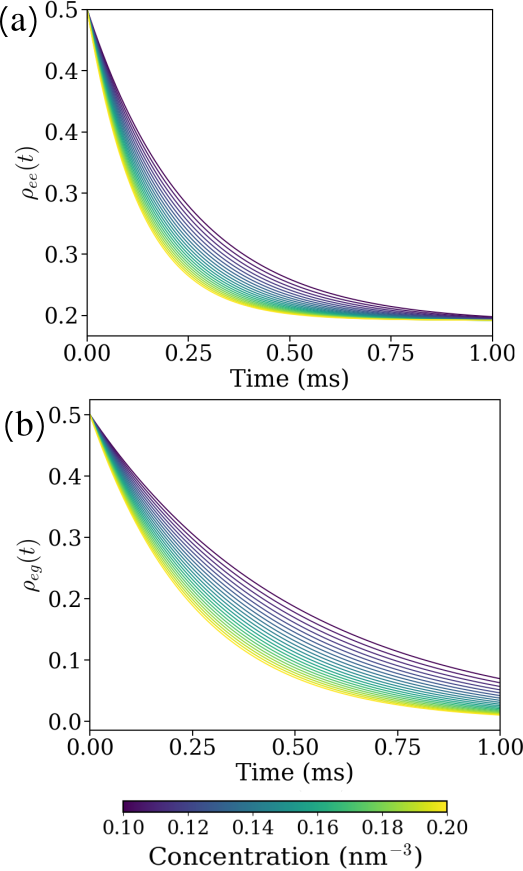} 
    \caption{\textbf{Time evolution of density matrix elements.} Evolution of (a) the excited-state population $\rho_{ee}(t)$ and (b) coherence $\rho_{eg}(t)$ at \(k_B T = 0.050\) eV. Increasing qubit concentration accelerates both population relaxation and coherence decay, indicating enhanced dissipation and decoherence. Parameters: $\omega_0 = 0.07$ eV, $R_{\min}=0.5$ nm, $R_{\max}=5$ nm, $R_0=1.5$ nm, and $\Gamma_{\text{SE}}=\gamma_0=100\ \mathrm{s}^{-1}$.}
    \label{dissipation}
\end{figure}
\begin{figure}
    \centering
    \includegraphics[width=0.45\textwidth]{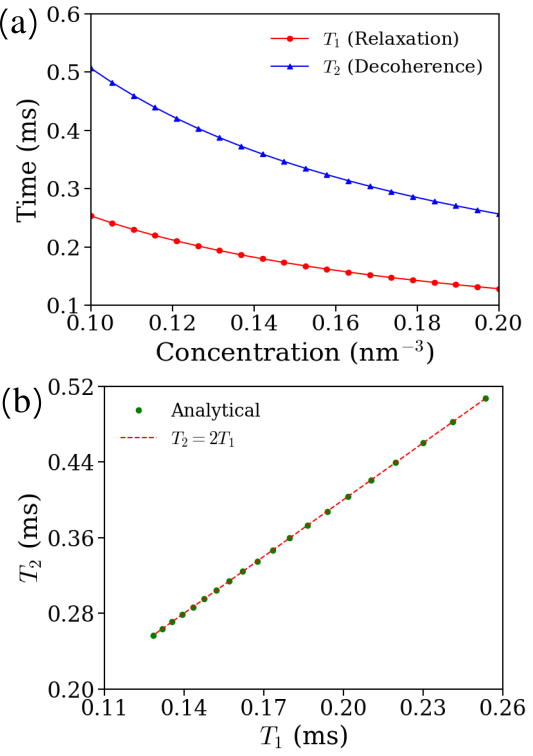} 
    \caption{\textbf{Relaxation and decoherence times.} The relaxation time $T_1$ and decoherence time $T_2$ are obtained at \(k_B T = 0.050\) eV, using $\omega_0=0.07$ eV, $R_{min}=0.5$ nm, $R_{max}=5$ nm, $R_0=1.5$ nm, and $\Gamma_{\text{SE}}=\gamma_0=100$ s$^{-1}$. (a) The top panel shows that both $T_1$ and $T_2$ decrease with increasing qubit concentration. (b) The bottom panel shows the relation between $T_2$ and $T_1$, confirming the expected relaxation-limited relation \(T_2 = 2T_1\) in the absence of pure dephasing.}
    \label{relaxation_decoherence}
\end{figure}

With these parameters established, we solve the MQMF-LME for the model multi-qubit system and analyze how qubit concentration controls the effective FRET-mediated transition rates and thereby influences the time evolution of the density matrix and the resulting relaxation and decoherence behavior. The diagonal elements describe the population dynamics of the quantum states, while the off-diagonal elements quantify the decay of quantum coherence. Since the MQMF-LME describes Markovian dynamics with time-independent transition rates, the elements of the density matrix exhibit exponential behavior. In particular, the excited-state population \(\rho_{ee}(t)\) relaxes exponentially toward its steady-state value, while the ground-state population \(\rho_{gg}(t)\) approaches its corresponding steady-state value as the system relaxes. The off-diagonal coherence \(\rho_{eg}(t)\) decays exponentially in amplitude. Figure~\ref{dissipation} shows the evolution of \(\rho_{ee}(t)\) and \(\rho_{eg}(t)\) over \(1\,\text{ms}\) timescale for qubit concentrations ranging from \(0.1\) to \(0.2\,\mathrm{nm}^{-3}\). The excited-state population \(\rho_{ee}(t)\) decreases exponentially toward a finite steady-state population of approximately \(0.2\), rather than decaying to zero, and reaches this steady state on a timescale of about \(1\,\text{ms}\). This finite population arises from bidirectional FRET-mediated excitation exchange between the system and bath: the system qubit can transfer excitation to the bath, while thermally populated bath qubits can transfer excitation back to the system. The balance between these processes is governed by the detailed-balance rates discussed in the Methods section. In contrast, the off-diagonal coherence \(\rho_{eg}(t)\) decays exponentially in amplitude and approaches zero over the same \(1\,\text{ms}\) timescale, indicating decoherence of the initial superposition state. Both population relaxation and coherence decay become faster as the concentration of bath qubits increases. The slowest decay is observed at the lowest concentration, whereas the fastest decay occurs at the highest concentration. This trend is expected because increasing the qubit concentration increases the number of available qubit--qubit interaction channels, thereby enhancing dissipation and decoherence. In Fig.~\ref{relaxation_decoherence}, we show the relaxation time \(T_1\) and the decoherence time \(T_2\) extracted from the simulated population and coherence dynamics. Both \(T_1\) and \(T_2\) decrease with increasing qubit concentration, reflecting the increase in the number of available qubit--qubit interaction channels as the bath density increases. Because the model system does not include an explicit pure-dephasing channel (\(1/T_\phi = 0\)), the extracted times follow the relaxation-limited relation \(T_2 = 2T_1\), consistent with the analytical result derived in the Methods section. 

When $1/f$ noise is incorporated into the MQMF-LME framework, an additional pure dephasing mechanism is introduced alongside the existing Lindblad dissipation channels. As derived in the Methods section, this noise contribution introduces an additional dephasing factor, $e^{-\chi_{1/f}(t)}$, in the off-diagonal elements of the density matrix. Consequently, the coherence of the qubit decays more rapidly compared to the noise-free case. Since the noise enters the Hamiltonian as a fluctuation in the qubit energy splitting, it affects only the phase evolution of the quantum state and does not alter the population dynamics governed by the diagonal elements of the density matrix. To examine the influence of $1/f$ noise on the decoherence dynamics, the noise amplitude and frequency cutoffs must be specified. In the present calculations, we use representative noise parameters, with amplitude $A=1.0\times10^{7} \mathrm{rad^2 s^{-2}}$, and a low-frequency cutoff of $1~\mathrm{Hz}$~\cite{ithier2005decoherence}, corresponding to an angular cutoff frequency $\omega_l=2\pi~\mathrm{rad \ s^{-1}}$. Here, $A$ controls the strength of the stochastic phase fluctuations and therefore determines the magnitude of the additional dephasing contribution. Larger values of $A$ produce a stronger suppression of the off-diagonal density matrix elements and a shorter decoherence time\cite{ithier2005decoherence,paladino20141}. The choice of $1~\mathrm{Hz}$ cutoff ensures that the simulation includes very slow fluctuations that remain nearly constant over the qubit evolution time. Such slow fluctuations are known to dominate phase accumulation in systems affected by $1/f$ noise and therefore provide the leading contribution to decoherence~\cite{paladino20141}. 

\begin{figure}
    \centering
    \includegraphics[width=0.4\textwidth]{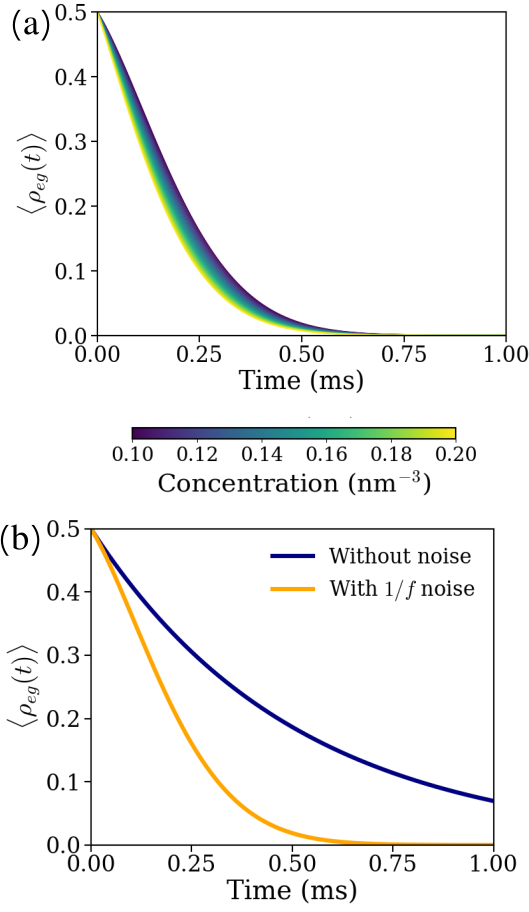} 
    \caption{\textbf{Time evolution of coherence in the presence of \(1/f\) noise.} (a) Decay of the ensemble-averaged coherence $\langle \rho_{eg}(t) \rangle$ at $k_BT=0.05$ eV for different qubit concentrations. Increasing qubit concentration accelerates coherence decay, indicating enhanced dissipation and decoherence. Parameters: $\omega_0 = 0.07$ eV, $R_{\min}=0.5$ nm, $R_{\max}=5$ nm, $R_0=1.5$ nm, and $\Gamma_{\text{SE}}=\gamma_0=100\ \mathrm{s}^{-1}$. The noise parameters are $A=1.0\times10^{7} \mathrm{rad^2 s^{-2}}$, and $\omega_l=2\pi~\mathrm{rad \ s^{-1}}$. (b) Comparison between the noise-free coherence dynamics from Fig.~\ref{dissipation} and the ensemble-averaged coherence $\langle \rho_{eg}(t) \rangle$ obtained after including $1/f$ noise, shown at a fixed qubit concentration of $0.1~\mathrm{nm^{-3}}$, while keeping all other parameters identical. The inclusion of $1/f$ noise introduces an additional pure dephasing channel, leading to faster coherence decay and a corresponding reduction in the decoherence time.
    }
    \label{fig:Dephasing}
\end{figure}

\begin{figure}
    \centering
    \includegraphics[width=0.45\textwidth]{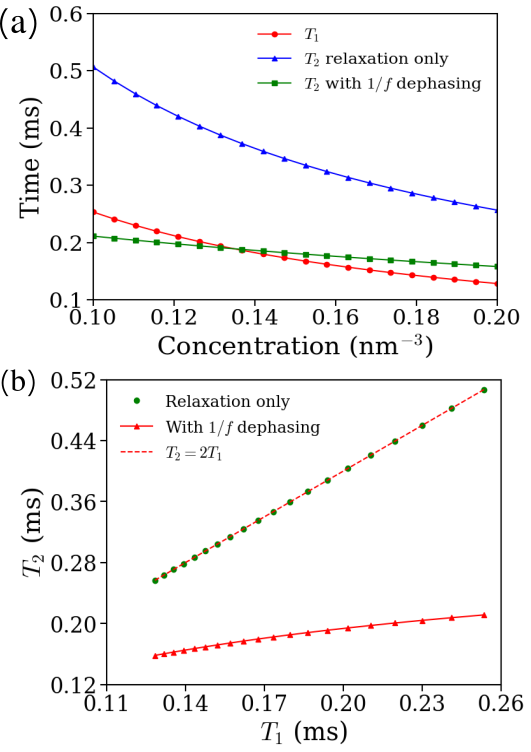} 
    \caption{\textbf{Relaxation and decoherence times with \(1/f\) noise.} The relaxation time $T_1$ and decoherence time $T_2$ are obtained at \(k_B T = 0.050\) eV, using $\omega_0=0.07$ eV, $R_{min}=0.5$ nm, $R_{max}=5$ nm, $R_0=1.5$ nm, and $\Gamma_{\text{SE}}=\gamma_0=100$ s$^{-1}$. (a) Comparison of \(T_1\), the noise-free \(T_2\), and the effective \(T_2\) obtained after including \(1/f\) noise. The addition of \(1/f\) noise significantly reduces the coherence time while leaving \(T_1\) largely unchanged. The noise-free \(T_1\) and \(T_2\) values are the same as those shown in Fig.~\ref{relaxation_decoherence} and are included here as a reference for comparison. (b) Relation between $T_2$ and $T_1$ at different qubit concentrations. In the absence of pure dephasing, the system follows the relaxation-limited relation $T_2=2T_1$. The presence of $1/f$ noise introduces an additional pure dephasing contribution ($1/T_\phi > 0$), causing the relation to deviate from this limit and resulting in $T_2 \leq 2T_1$.
}
    \label{fig:Dephasing-effect}
\end{figure}

Figure~\ref{fig:Dephasing} shows the effect of this additional dephasing channel on the coherence dynamics. Figure~\ref{fig:Dephasing}(a) shows the time evolution of the ensemble-averaged coherence $\langle \rho_{eg} \rangle$ for different qubit concentrations in the presence of \(1/f\) noise. As in the noise-free case, increasing qubit concentration accelerates coherence decay because it increases the number of available qubit--qubit interaction channels. Figure~\ref{fig:Dephasing}(b) compares the coherence dynamics at a fixed concentration of 0.1 nm$^{-3}$ with and without $1/f$ noise. The noise-free MQMF-LME result shows the deterministic coherence \(\rho_{eg}(t)\)(blue curve), whereas the with-noise result shows the ensemble-averaged coherence \(\langle\rho_{eg}(t)\rangle\) (orange curve). The inclusion of \(1/f\) noise introduces an additional dephasing contribution and accelerates the decay of the coherence. This comparison demonstrates that environmental noise causes loss of phase coherence beyond that produced by the Lindblad dissipative channels alone. The impact of this additional dephasing on the characteristic timescales is summarized in Fig.~\ref{fig:Dephasing-effect}. Figure~\ref{fig:Dephasing-effect}(a) compares the relaxation time $T_1$ (red), the decoherence time  $T_2$ obtained without noise (blue), and the effective decoherence time obtained after including $1/f$ noise (green). While  $T_1$ remains essentially unchanged, the inclusion of noise significantly reduces $T_2$, demonstrating that the noise primarily affects the phase coherence of the qubit rather than the population dynamics. This behavior is further reflected in Fig.~\ref{fig:Dephasing-effect}(b), which examines the relationship between $T_1$ and $T_2$. In the absence of pure dephasing, coherence decay is governed solely by relaxation processes (green symbols), yielding the relaxation-limited relation $T_2 = 2T_1$ (red dashed line). However, the introduction of $1/f$ noise creates an additional pure dephasing contribution. When this additional decay is represented by an effective pure-dephasing time $T_{\phi}$, the decoherence rate becomes $\frac{1}{2T_1}+\frac{1}{T_\phi}$. As a result, the coherence time satisfies $T_2 \leq 2T_1$ (red triangles and solid line), with equality recovered only in the absence of pure dephasing. The deviation observed in Fig.~\ref{fig:Dephasing-effect}(b) therefore indicates the additional dephasing introduced by $1/f$ noise.

\subsection{Application to Er$^{3+}$-Doped CeO$_2$}

\begin{figure}
    \centering
    \includegraphics[width=0.5\textwidth]{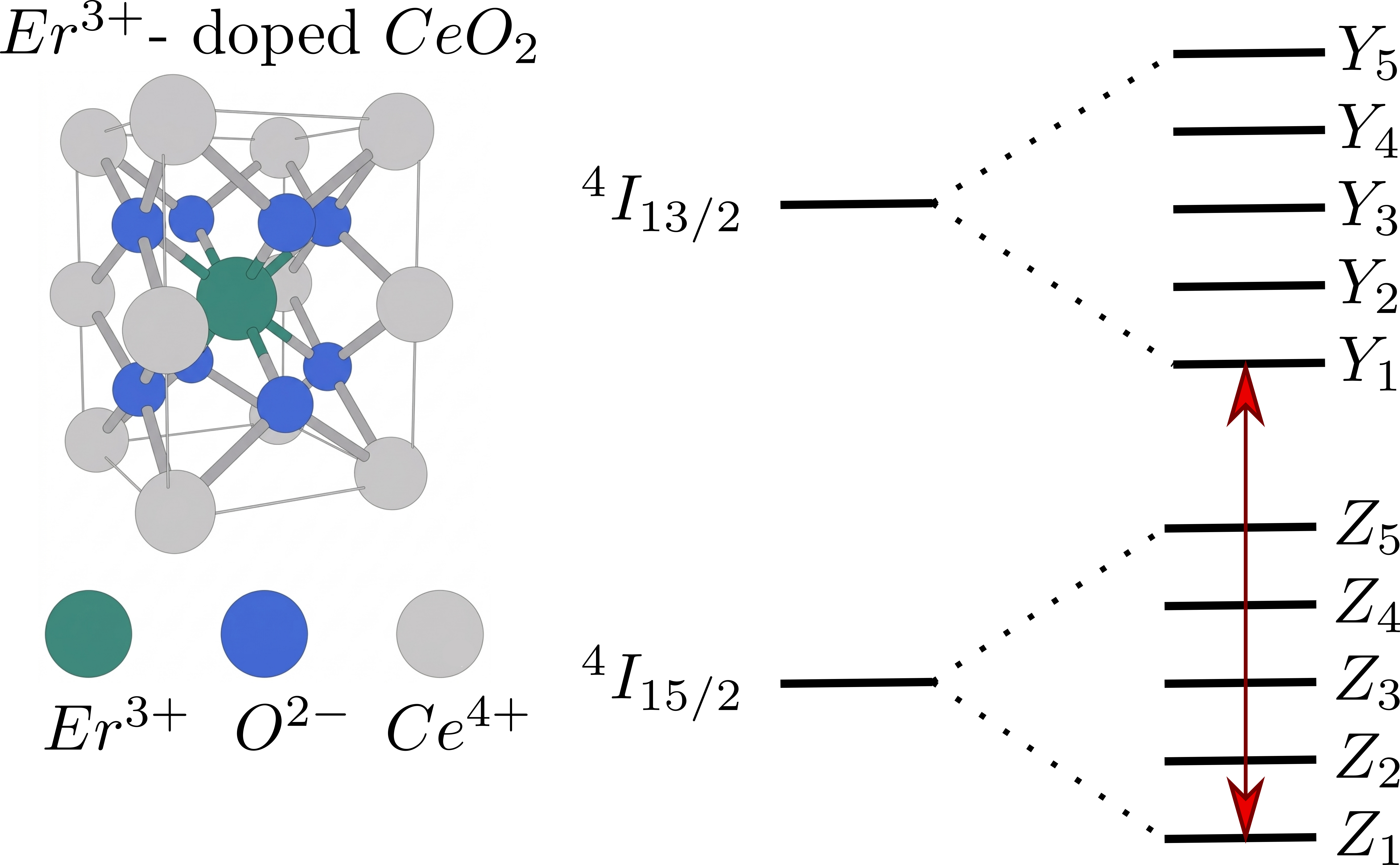} 
    \caption{\textbf{Optical transition of Er$^{3+}$-doped Ce$O_2$.} Er$^{3+}$ has a$4f^{11}$ electronic configuration, with the $^{4}I_{15/2}$ and $^{4}I_{13/2}$ manifolds corresponding to the ground and first excited states, respectively. In the presence of the crystal field, these manifolds split into five nondegenerate crystal-field levels labeled $Z_1$ to $Z_5$ and $Y_1$ to $Y_5$, respectively~\cite{zhang2024optical, wong2024coherent}. The $Y_1$--$Z_1$ transition lies in the telecom C-band and can be addressed optically.}
    \label{fig:Er-doped-CeO2}
\end{figure}

We next apply the MQMF-LME framework to Er$^{3+}$-doped CeO$_2$, a rare-earth-ion platform of interest for fiber-based quantum networks~\cite{wong2024coherent,grant2024optical}. CeO$_2$ crystallizes in a cubic fluorite structure, providing high symmetry, wide bandgap transparency, and stable substitutional sites for Er$^{3+}$ ions. The localized \(4f\) orbitals of Er$^{3+}$ are shielded by the outer $5s$ and $5p$ orbitals, resulting in sharp optical transitions that are relatively insensitive to lattice vibrations. The transition between the lowest crystal-field levels of the $^{4}I_{13/2}$ excited manifold and the $^{4}I_{15/2}$ ground manifold has an energy difference of approximately 0.8 eV, placing it in the telecom band~\cite{grant2024optical, zhang2024optical}. The relevant crystal-field level structure and optically addressable $Y_1$--$Z_1$ transition are shown in Fig.~\ref{fig:Er-doped-CeO2}. This makes Er$^{3+}$-doped CeO$_2$ a promising solid-state qubit platform. Grant \textit{et al.} used time-resolved photoluminescence spectroscopy to examine the relaxation dynamics of Er\textsuperscript{3+} ions in CeO\textsubscript{2}~\cite{grant2024optical}. 
At low Er\textsuperscript{3+} concentrations, the decay traces were nearly exponential, with relaxation times close to the intrinsic radiative lifetime of the $^{4}I_{13/2}$ level. However, as the Er\textsuperscript{3+} concentration increased, the relaxation time decreased significantly and the decay curves became increasingly non-exponential. This concentration-quenching behavior was attributed to enhanced Er\textsuperscript{3+}--Er\textsuperscript{3+} interactions~\cite{grant2024optical}, including cross-relaxation and resonant excitation transfer, which introduce additional nonradiative pathways for the depopulation of the $^{4}I_{13/2}$ state. This experimentally observed concentration dependence provides a direct test case for the MQMF-LME framework.

\begin{figure}
    \centering
    \includegraphics[width=0.45\textwidth]{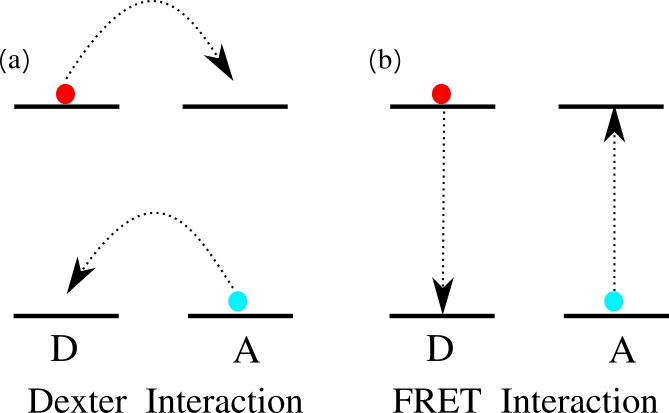} 
    \caption{\textbf{Excitation transfer mechanisms}. (a) Schematic representation of Dexter-type and (b) FRET-mediated excitation transfer. \(D\) and \(A\) denote the donor and acceptor, respectively. In the Dexter mechanism, excitation transfer occurs through a simultaneous electron-exchange process: the donor de-excites while the acceptor is promoted to its excited state. Because this process requires wavefunction overlap between donor and acceptor, Dexter transfer is short ranged and typically operates only at very small separations. In contrast, in FRET, excitation transfer occurs through long-range nonradiative dipole--dipole coupling: the donor de-excites while the acceptor is promoted to its excited state. Because this process does not require electron exchange or wavefunction overlap, FRET can operate over larger donor--acceptor separations than Dexter transfer.}
    \label{etransfer}
\end{figure}

To identify the dominant Er\textsuperscript{3+}--Er\textsuperscript{3+} interaction mechanism, we compare two limiting nonradiative excitation-transfer processes: short-range Dexter-type exchange and long-range FRET-mediated dipole--dipole transfer, as illustrated in Fig.~\ref{etransfer}. FRET is a nonradiative dipole--dipole excitation-transfer process and is typically effective over donor--acceptor separations of approximately 1--10 nm~\cite{FRET,FRET2, lakowicz2006principles}. Dexter transfer, by contrast, requires direct orbital overlap and is therefore restricted to much shorter separations, typically below 1 nm~\cite{Dexter,skourtis2016dexter,lakowicz2006principles}. We first consider Dexter-type transfer. In this model, the excitation transfer rate between a central Er$^{3+}$ ion, labeled 0, and the $i^\text{th}$ bath ion is 
\begin{equation*}
\gamma_{0i}= \gamma_0 \exp \left [\frac{2R_0}{L}\left(1-\frac{R_{0i}}{R_0}\right) \right],
\end{equation*} 
where $\gamma_0$ is the reference transfer rate, $L$ is the effective orbital decay length, and $R_{0i}$ is the distance between Er\textsuperscript{3+} ions. Because Dexter transfer is short ranged, we restrict the interaction distance to $R_\text{min}=0.5$ nm and $R_\text{max}=1$ nm. Since the characteristic Dexter distance $R_0$ is not known a priori for Er$^{3+}$-doped CeO$_2$, we vary $R_0$ from 0.5 to 1 nm. The energy splitting is set to 0.81 eV, corresponding to the Er$^{3+}$ telecom transition, and bath temperature is set to $T=4$K ($k_BT=0.00034$ eV), matching the experimental conditions. The intrinsic relaxation rate is set to $\Gamma_{\text{SE}}= 285$ s$^{-1}$, corresponding to a single-ion lifetime of approximately 3.5 ms, as measured in the low-concentration limit~\cite{grant2024optical}. 

\begin{figure}
    \centering
    \includegraphics[width=0.4\textwidth]{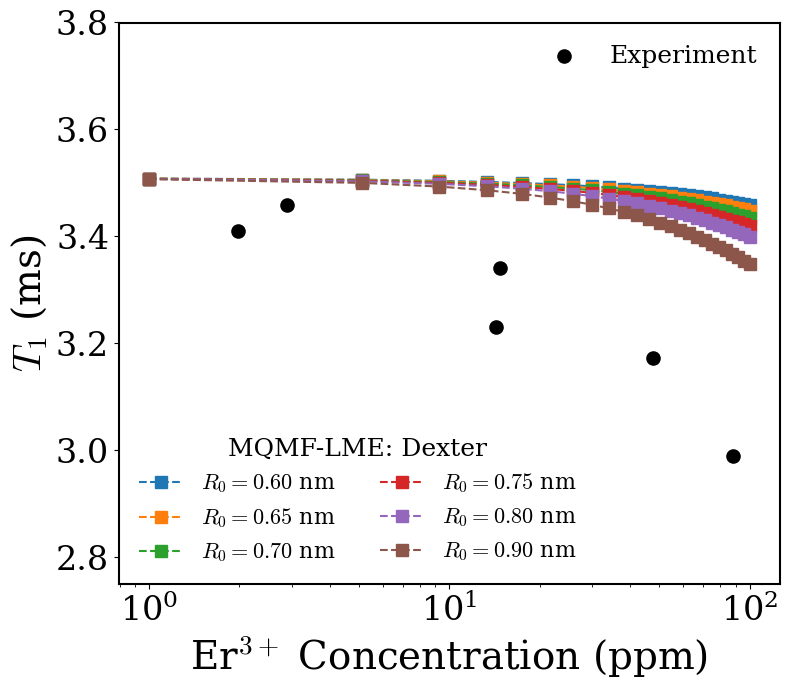} 
    \caption{\textbf{Dexter-based MQMF-LME simulation for Er$^{3+}$-doped CeO$_2$.} Simulated relaxation time \(T_1\) as a function of Er$^{3+}$ concentration for Dexter-type excitation transfer. The experimental data, extracted from Grant \textit{et al.}~\cite{grant2024optical}, are shown as black circles, and the MQMF-LME Dexter simulations are shown as square symbols with dashed lines. The characteristic Dexter distance  $R_0$ is varied from 0.5 to 1 nm, while $R_{\text{min}} = 0.5$ nm, $R_{\text{max}} = 1$ nm, $\gamma_0 = \Gamma_{\text{SE}}= 285$ s$^{-1}$, and $T = 4$ K are held fixed. Larger $R_0$ values increase the transfer rate, but even the strongest Dexter case fails to reproduce the experimental concentration dependence, indicating that Dexter-type exchange is not the dominant relaxation mechanism.}
    \label{fig:Dexter}
\end{figure}

Figure~\ref{fig:Dexter} compares the experimentally measured relaxation times with MQMF-LME simulations based on Dexter-type transfer. Across the full range of $R_0$ values considered, the simulated relaxation time decreases much more slowly with increasing concentration than observed experimentally. This mismatch indicates that Dexter-type exchange alone cannot account for the concentration-dependent relaxation observed in Er$^{3+}$-doped CeO$_2$. This conclusion is consistent with the microscopic requirements of Dexter transfer: the process relies on electron exchange and therefore requires strong orbital overlap between neighboring ions. Er$^{3+}$ has a $4f^{11}$ electronic configuration~\cite{wang2017er3+}, and the $4f$ orbitals are strongly shielded by the outer $5s$, $5p$, and $6s$ electrons~\cite{abragam2012electron,rat,eliseeva2010lanthanide}. As a result, the spatial overlap between neighboring Er$^{3+}$ ions is expected to be weak, making Dexter-type exchange negligible. We therefore next consider the long-range FRET-mediated interactions.

For FRET-mediated excitation transfer, we use the distance-dependent rate introduced in
Eq.~\ref{eq:gamma-FRET}. Because FRET is effective over nanometer-scale donor--acceptor separations~\cite{lakowicz2006principles}, we choose $R_{max} = 5$ nm as the upper cutoff. We set $R_{min}=0.5$ nm as a closest-approach cutoff to exclude unphysically short Er$^{3+}$-Er$^{3+}$ separations in the continuum average. The characteristic FRET distance $R_0$ then sets the overall interaction strength. This parameter depends on both the optical properties of the Er$^{3+}$ transition and the dielectric environment of the CeO$_2$ host, and can be estimated from the following expression~\cite{lakowicz2006principles}:
\begin{equation}
    R_0^6 = \frac{9000 (ln10)\kappa^2 Q_D}{128\pi^5N_A n^4}J,
    \label{eq:critical transfer}
\end{equation}
where $Q_D$ is the quantum yield of the donor, $\kappa^2$ is the dipole-orientation factor, $n$ is the refractive index of the host material, $N_A$ is Avogadro's number and $J$ is the  spectral overlap integral. Here, $N_A$ is used for Avogadro's number to avoid confusion with the number of bath qubits, $N$, introduced in the MQMF-LME framework. For randomly oriented dipoles, we take $\kappa^2 = \frac{2}{3}$~\cite{lakowicz2006principles}. The refractive index of $\mathrm{CeO_2}$ is taken to be $n=2.3$. The spectral overlap integral $J$ is defined as:
\begin{equation}
    J =\int_o ^\infty F_D(\lambda)\epsilon_A(\lambda) \lambda^4 d\lambda, 
    \label{eq:Spectral overlap integral}
\end{equation}
where $F_D(\lambda)$ is the normalized donor emission spectrum, $\epsilon_A(\lambda)$ is the molar absorption coefficient of the acceptor, and $\lambda$ is the wavelength. The donor emission spectrum is normalized such that $\int_0^\infty F_D(\lambda)\, d\lambda = 1.$ To obtain an approximate estimate for Er$^{3+}$-doped CeO$_2$, we treat the donor emission as sharply peaked at the telecom transition wavelength, $\lambda_0 \approx 1530$ nm, corresponding to the $\sim0.81$ eV Er$^{3+}$ transition, and approximate it as $F_D(\lambda) = \delta(\lambda - \lambda_0)$. 
With this approximation, the spectral overlap integral reduces to 
\begin{equation}
    J =\int_0^\infty \delta (\lambda-\lambda_0) \epsilon_A(\lambda) \lambda^4 d\lambda\\
    = \epsilon_A(\lambda_0)\lambda_0^4.
    \label{eq:Spectral Overlap integral calculation}
\end{equation}
The spectral overlap integral $J$ quantifies the overlap between the donor emission spectrum and the acceptor absorption spectrum. A larger $J$ increases the characteristic FRET distance $R_0$, thereby enhancing the effective donor--acceptor excitation-transfer rate. The molar absorption coefficient $\epsilon_A$ is related to the absorption cross section $\sigma$ using~\cite{lakowicz2006principles}:
\begin{equation}
    \epsilon_A = \frac{\sigma(\lambda) N_A}{1000 \ln 10}.
    \label{eq:Molar excitation coefficeint}
\end{equation}
The absorption cross section of an acceptor quantifies the likelihood that it absorbs light at a specific wavelength. This quantity can be influenced by defects, impurities, local crystal-field distortions, and near-field effects in a host material~\cite{Absorption, Absorption2}. For Er$^3+$, reported absorption cross sections span approximately $10^{-19}-10^{-22}$ cm $^2$~\cite{barnes2002absorption, barnes1990absorption}, depending on the host and local environment~\cite{Absorption1, Absorption3}.
The donor quantum yield is also difficult to determine precisely for Er$^{3+}$-doped CeO$_2$ because it depends on local-field effects and competing non-radiative decay pathways. Reported values for $\mathrm{Er^{3+}}$ in related host materials typically fall in the range $Q_D=0.08-0.3$~\cite{tatar2013influence}. 

\begin{figure}
    \centering
    \includegraphics[width=0.5\textwidth]{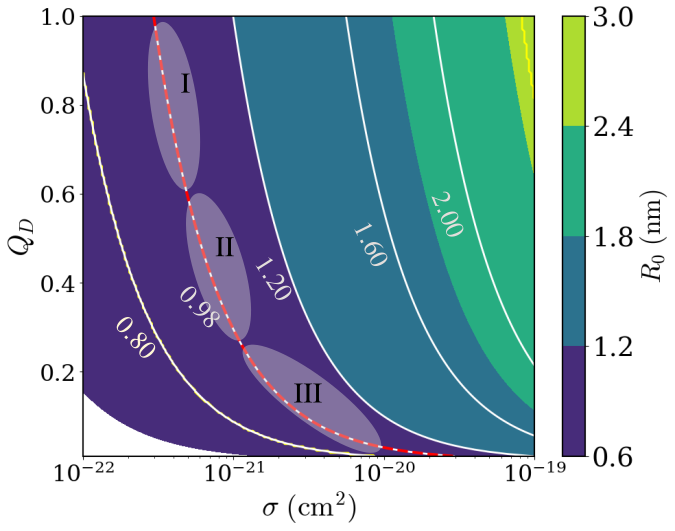} 
    \caption{\textbf{Phase diagram of the characteristic FRET distance $R_0$.} Calculated $R_0$ values as a function of donor quantum yield $Q_D$ and absorption cross section $\sigma$ for Er$^{3+}$--Er$^{3+}$ FRET in CeO$_2$. White contour lines indicate representative $R_0$ values in the range 0.8--2.5 nm. The diagram shows that the best-fit value $R_0=0.98$ nm lies within the physically plausible range of $Q_D$ and $\sigma$.}
    \label{fig:Phase-diagram}
\end{figure}

In Eq.~\ref{eq:gamma-FRET}, $R_0$ denotes the characteristic F\"orster distance at which the FRET transfer efficiency reaches 50\%. Because $R_0$ depends on both the donor quantum yield $Q_D$ and the spectral overlap integral $J$, uncertainties in $Q_D$ and the absorption cross section $\sigma$ lead to uncertainty in the effective FRET interaction strength. We therefore vary $Q_D$ and $\sigma$ to estimate the physically reasonable range of $R_0$, as shown in Fig.~\ref{fig:Phase-diagram}.
The phase diagram indicates that $R_0$ values between 0.8 and 2.5 nm are physically reasonable for Er$^{3+}$-doped CeO$_2$. Higher values of $Q_D$ and $\sigma$ increase $R_0$, reflecting fewer competing nonradiative decay pathways and stronger donor--acceptor spectral overlap. Conversely, lower absorption cross sections reduce the spectral overlap and decrease the effective FRET distance. Importantly, the same value of $R_0$ can arise from different combinations of $Q_D$ and $\sigma$, indicating that $R_0$ should be interpreted as an effective interaction parameter that incorporates multiple local-environment effects. Beyond the present system, this type of phase diagram provides a practical way to connect uncertain optical parameters to an effective interaction length, which can guide the selection of physically reasonable $R_0$ values in other rare-earth-ion or defect-based qubit materials. A more detailed discussion of the $Q_D-\sigma-R_0$ relationship is provided in Appendix~\ref{app:fret-distance}.

\begin{figure}
    \centering
    \includegraphics[width=0.4\textwidth]{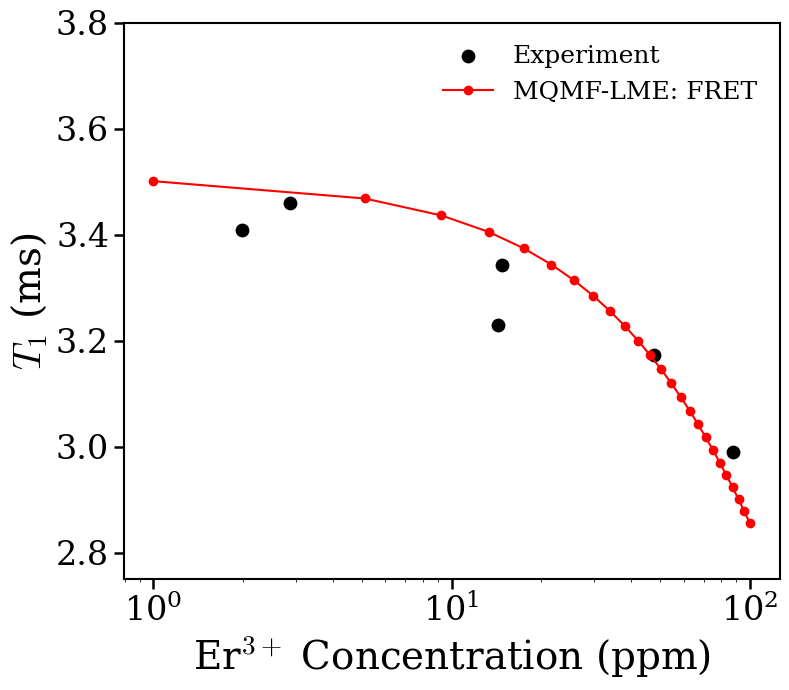} 
    \caption{\textbf{FRET-based MQMF-LME Simulation for Er$^{3+}$-doped CeO$_2$.} Simulated relaxation time $T_1$ for FRET-mediated excitation transfer with $R_0$= 0.98 nm. The experimental data, extracted from Grant \textit{et al.}~\cite{grant2024optical}, are shown as black circles, and the MQMF-LME FRET result is shown as red symbols and a line. Simulation parameters are $R_{\text{min}} = 0.5$ nm, $R_{\text{max}} = 5$ nm, $\gamma_0 = \Gamma_{\text{SE}}= 285$ s$^{-1}$, $T = 4$ K. The MQMF-LME result shows close agreement with the experimental measurements, indicating that FRET-mediated excitation transfer is the dominant interaction mechanism in Er$^{3+}$-doped CeO$_2$.}
    \label{fig:expt-comparison}
\end{figure}

Within the physically reasonable range identified in Fig.~\ref{fig:Phase-diagram}, the MQMF-LME simulations show the closest agreement with experiment for $R_0 = 0.98 $ nm. Figure~\ref{fig:expt-comparison} compares the resulting MQMF-LME prediction with the experimentally measured relaxation time as a function of Er$^{3+}$ concentration.
Figures~\ref{fig:Dexter} and~\ref{fig:expt-comparison} show that the Dexter- and FRET-based models produce similar relaxation behavior at low Er$^{3+}$ concentrations, where the ions are far apart and inter-ion interactions are weak. As the concentration increases, however, the two models diverge. The FRET-based model reproduces the experimentally observed decrease in $T_1$, whereas the Dexter-based model substantially underestimates the concentration dependence. This comparison identifies long-range FRET-mediated excitation transfer, rather than short-range Dexter exchange, as the dominant mechanism responsible for concentration-dependent relaxation in Er$^{3+}$-doped CeO$_2$. This distinction has important implications for engineering rare-earth-ion qubit materials. If relaxation is dominated by FRET-mediated dipole--dipole coupling, decoherence and concentration quenching can be mitigated by controlling dopant concentration, increasing average ion spacing, or selecting host materials that reduce effective dipolar coupling through dielectric screening or reduced spectral overlap. If Dexter exchange were dominant, mitigation would instead require suppressing direct wavefunction overlap, which is a substantially more restrictive materials-design constraint.

\section{Conclusion}

Decoherence in multi-qubit systems generally increases with qubit concentration, indicating that qubit--qubit interactions can play a major role in limiting qubit performance. However, the microscopic interaction mechanisms responsible for concentration-dependent relaxation and decoherence are often difficult to identify. In this work, we developed a multi-qubit mean-field Lindblad master equation (MQMF-LME) framework to address this challenge. The framework captures qubit--qubit interactions through effective bath-induced transition rates and provides a tractable description of relaxation and decoherence in interacting multi-qubit systems. Within the MQMF-LME framework, the system qubit experiences three primary dissipative processes: intrinsic single-qubit relaxation, excitation transfer from the system qubit to surrounding bath qubits, and excitation transfer from thermally populated bath qubits back to the system qubit. This formulation connects relaxation and decoherence dynamics to qubit concentration, spatial distribution, interaction range, and bath occupation. It therefore provides a physically interpretable way to relate microscopic qubit--qubit interaction mechanisms to experimentally measurable timescales. A key feature of the present work is that the MQMF-LME framework provides both analytical and numerical access to multi-qubit decoherence. The analytical solutions clarify how the population and coherence dynamics depend on the upward and downward transition rates, identify the steady-state populations, and establish the relaxation-limited relation $T_2=2T_1$ in the absence of pure dephasing. These closed-form results provide physical interpretability and serve as internal consistency checks for the numerical implementation. The simulations then extend this analytical foundation beyond the closed-form baseline by enabling direct analysis of concentration-dependent dynamics, noise-induced dephasing, and material-specific excitation-transfer mechanisms.

For a model multi-qubit system with FRET-mediated dipole--dipole excitation exchange, the MQMF-LME simulations show that both the relaxation time $T_1$ and the decoherence time $T_2$ decrease with increasing qubit concentration. This trend arises because increasing the bath density increases the number of available qubit--qubit interaction channels. In the absence of an explicit pure-dephasing mechanism, the model follows the relaxation-limited relation $T_2 = 2T_1$, consistent with the analytical solution of the MQMF-LME. We further extended the framework to include noise-induced pure dephasing by introducing stochastic fluctuations in the qubit energy splitting. For $1/f$ noise, the off-diagonal density-matrix elements acquire an additional non-exponential dephasing envelope, $e^{-\chi_{1/f}(t)}$. This additional noise channel reduces the effective decoherence time $T_2$ while leaving the population relaxation time $T_1$ unchanged, demonstrating that environmental fluctuations can produce coherence loss beyond that caused by relaxation alone.

To validate our framework in a realistic materials platform, we applied it to Er$^{3+}$-doped CeO$_2$, a rare-earth-ion system with an optically addressable telecom-band transition. Er$^{3+}$ ions exhibit sharp intra-$4f$ optical transitions, and the $Y_1-Z_1$ transition can be addressed in the telecom C-band, making this system relevant for optical quantum memory. Experimental observations showed that the relaxation time of this transition decreases with increasing Er$^{3+}$ concentration, indicating enhanced interactions between Er$^{3+}$ ions. To explore these interactions, we compared two limiting excitation-transfer mechanisms: short-range Dexter-type exchange and long-range FRET-mediated dipole--dipole transfer. Dexter-type transfer, which requires direct orbital overlap, failed to reproduce the experimentally observed concentration dependence. This result is consistent with the localized and shielded nature of the Er$^{3+}$ $4f$ orbitals, which strongly suppresses wavefunction overlap between neighboring ions. In contrast, the FRET-based MQMF-LME model reproduced the experimental relaxation trend using a physically reasonable characteristic FRET distance. This comparison identifies long-range FRET-mediated excitation transfer as the dominant mechanism responsible for concentration-dependent relaxation in Er$^{3+}$-doped CeO$_2$. 
 
More broadly, the MQMF-LME framework provides a modular and generalizable approach for connecting microscopic qubit--qubit interactions to experimentally measurable relaxation and decoherence times. Because the interaction mechanism enters through physically interpretable transition rates, Lindblad operators, and stochastic Hamiltonian terms, the framework can be adapted to different solid-state qubit platforms, including rare-earth ions, color centers, defect qubits, and other doped quantum materials. It can also incorporate multiple decoherence pathways, such as dipole--dipole excitation transfer, exchange-mediated transfer, spontaneous emission, charge noise, magnetic-field noise, defect-induced fluctuations, and host-mediated pure dephasing. By linking material parameters, dopant concentration, spatial distribution, and environmental noise to $T_1$ and $T_2$, the MQMF-LME framework provides a practical route for identifying dominant microscopic loss channels and designing solid-state qubit materials with reduced concentration-induced decoherence.

\section{Acknowledgements}

We gratefully acknowledge Prof. Supratik Guha, Dr. Greg Grant, and Dr. Swarbnabha Chattaraj from the University of Chicago and Argonne National Laboratory for helpful discussions and guidance related to Er$^{3+}$-doped CeO$_2$. This work utilized the Alpine high performance computing resource at the University of Colorado Boulder. Alpine is jointly funded by the University of Colorado Boulder, the University of Colorado Anschutz, and Colorado State University and with support from NSF grants OAC-2201538 and OAC-2322260.

\setcounter{secnumdepth}{1}

\appendix

\section{Derivation of the Lindblad Master Equation under the Born--Markov and Rotating Wave Approximations}
\label{app:LME}

In the Methods section, we introduced the system--bath model and the corresponding Hamiltonians. The total dynamics are governed by the von Neumann equation in Eq.~\ref{eq:VonNeuman}. Here, we summarize how the Lindblad master equation follows from this starting point under the standard Born, Markov, and secular approximations. To focus on the effect of the system--bath interaction, we work in the interaction picture, where the time evolution of the total density matrix is governed by
\begin{equation}
    \frac{d \rho_T^I(t)}{dt} = -i \alpha \left[ H_{SB}^I(t),\ \rho_T^I(t) \right].
    \label{eq:interaction picture}
\end{equation}
Here, $\rho_T^I(t)$ and $H_{SB}^I(t)$ are the total density matrix and the interaction Hamiltonian, respectively, in the interaction picture. Reference~\cite{galindo2012quantum} provides a detailed discussion of the Schr\"odinger, Heisenberg, and Interaction pictures.
Integrating Eq.~\ref{eq:interaction picture} gives 
\begin{equation}
    \rho_T^I(t) = \rho_T^I(0) - i\alpha \int_0^td\tau\left[ H_{SB}^I(\tau), \ \rho_T^I(\tau) \right]
    \label{eq:integral}
\end{equation}
Substituting Eq.~\ref{eq:integral} back into Eq.~\ref{eq:interaction picture} and taking the trace over the bath degrees of freedom yields
\begin{equation}
    \frac{d \rho_S^I(t)}{dt} = -\alpha^2\int_0^t d\tau  Tr_B\left[ H_{SB}^I(t),\left[ H_{SB}^I(\tau), \rho_T^I(\tau) \right]\right],
    \label{eq:redfield1}
\end{equation}
where we assume $\mathrm{Tr}_B \left[ H_{SB}^I(t), \rho_T^I(0) \right] = 0$. This condition is satisfied when the bath operators have zero mean, or equivalently after redefining the system Hamiltonian to absorb any nonzero bath average.  
To simplify Eq.~\ref{eq:redfield1}, we invoke the Born approximation, which assumes weak system--bath coupling and a bath that is sufficiently large that its state is negligibly affected by the system. Under this approximation, the total density matrix can be factorized as
\begin{equation}
    \rho_T^I(t) \approx \rho_S^I(t) \otimes \rho_B.
    \label{eq:Bornmarkov}
\end{equation}
Here, $\rho_B$ is the stationary equilibrium state of the bath. Substituting Eq.~\ref{eq:Bornmarkov} into Eq.~\ref{eq:redfield1} gives a time-nonlocal integro-differential equation for the reduced density matrix of the system,
\begin{equation}
\begin{aligned}
\frac{d \rho_S^I(t)}{dt}
&=
-\alpha^2\int_0^t d\tau \,
\mathrm{Tr}_B
\Big[
H_{SB}^I(t),
\\
&\qquad
\Big[
H_{SB}^I(\tau),
\rho_S^I(\tau)\otimes\rho_B
\Big]
\Big].
\end{aligned}
\label{eq:redfield2}
\end{equation}
The equation is not yet Markovian because the evolution of the reduced density matrix at time $t$ depends on its earlier values, $\rho_S^I(\tau)$. To simplify the equation, we employ the Markov approximation, in which $\rho_S^I(\tau)$ is replaced by $\rho_S^I(t)$. This approximation is valid when the bath correlation time is much shorter than the characteristic relaxation timescale of the system. Under this approximation, the time evolution of the system density matrix at time $t$ depends only on its current state,
\begin{equation}
\begin{aligned}
\frac{d \rho_S^I(t)}{dt}
&=
-\alpha^2\int_0^t d\tau \,
\mathrm{Tr}_B
\Big[
H_{SB}^I(t),
\\
&\qquad
\Big[
H_{SB}^I(\tau),
\rho_S^I(t)\otimes\rho_B
\Big]
\Big].
\end{aligned}
\label{eq:Final_Redfield}
\end{equation}
If the bath correlation function decays rapidly, the upper limit of the integral can be extended to infinity. After changing the integration variable from $\tau$ to $t-\tau$, the Born--Markov Redfield equation becomes~\cite{redfield1957theory, carver1981use}
\begin{equation}
\begin{split}
\frac{d \rho_S^I(t)}{dt} = -\alpha^2 \int_0^\infty d\tau \, 
& \, \mathrm{Tr}_B \Big[ H_{SB}^I(t),  \\
& \quad \left[ H_{SB}^I(t-\tau), \rho_S^I(t) \otimes \rho_B \right] \Big].
\end{split}
\end{equation}
This equation is time local, but it is not yet in Lindblad form and does not, in general, guarantee complete positivity of the reduced density matrix. To obtain a completely positive Markovian master equation, we apply the secular approximation, also referred to as the rotating wave approximation~\cite{RWA}. We write the system--bath interaction Hamiltonian as $H_{SB} = \sum_i S_i \otimes B_i$, where $S_i$ and $B_i$ act on the Hilbert spaces of the system and bath, respectively. The secular approximation is most conveniently performed by decomposing each system operator into eigenoperators of the system Hamiltonian $H_S$,
\begin{equation}
    S_i = \sum_{\omega} S_i(\omega).
\end{equation}
$S_i(\omega)$ must satisfy
\begin{equation}
    \left[ H_S, S_i(\omega)\right] = -\omega S_i(\omega),\quad  [ H_S, S_i^\dagger(\omega)] = \omega S_i^\dagger(\omega).
\end{equation}
Equivalently, $S_i(\omega)$ connects energy eigenstates of $H_S$ whose energy difference is $\omega$. In the interaction picture, 
\begin{equation}
    S_i^I(t) = e^{iH_S t} S_i e^{-iH_S t}. 
\end{equation}
The bath operators evolve as
\begin{equation}
    B_i^I(t) = e^{iH_Bt} B_i e^{-iH_Bt}. 
\end{equation}

We assume that the bath operator satisfies $\langle B_i^I(t)\rangle = Tr[B_i^I(t)\rho_B]=0$. If this condition is not initially satisfied, the nonzero mean can be absorbed into a renormalization of the system Hamiltonian. Using the eigenoperator decomposition in the Born--Markov Redfield equation, the master equation can be written as
\begin{equation}
     \frac{d \rho_S^I(t)}{dt} = \sum_{\omega,\omega'}\sum_{i,j}e^{i(\omega'-\omega)t}\Gamma_{ij}(\omega) [S_j(\omega)\rho_S^I,S_i^{\dagger}(\omega')]+h.c.,
     \label{eq:secular}
\end{equation}
where h.c. denotes the Hermitian Conjugate. The bath response is contained in the one-sided Fourier transform of the bath correlation function,
\begin{equation}
     \Gamma_{ij}(\omega) =\int_0^\infty d\tau e^{i\omega\tau}\langle B_i^{I\dagger}(\tau)B_j^{I}(0)\rangle
\end{equation}
Here, the two indices $i$ and $j$ are retained because different bath operators can be correlated. Since the bath is assumed to be in thermal equilibrium, [$H_B, \rho_B$]=0, the bath correlation functions depend only on time differences. As a result, $\Gamma_{ij}(\omega)$ is time independent. The secular approximation removes rapidly oscillating terms in the master equation. Terms with $\omega \neq \omega^\prime$ oscillate with frequency $|\omega-\omega^\prime|$ and average to zero when their oscillation period is much shorter than the relaxation timescale of the system. Therefore, only terms with $\omega'=\omega$ are retained. This yields
\begin{equation}
     \frac{d \rho_S^I(t)}{dt} = \sum_{\omega}\sum_{i,j}\Gamma_{ij}(\omega) [S_j(\omega)\rho_S^I(t),S_i^{\dagger}(\omega)]+h.c.
\end{equation}
To separate the dynamics into Hamiltonian and dissipative contributions, we decompose $\Gamma_{ij}(\omega)$ into real and imaginary parts,
\begin{equation}
    \Gamma_{ij}(\omega) =\frac{1}{2} G_{ij}(\omega) + i\Lambda_{ij}(\omega)
\end{equation}
\begin{equation}
\begin{split}
    G_{ij}(\omega) 
    &= \Gamma_{ij}(\omega)+\Gamma_{ji}^*(\omega) \\
    &= \int_{-\infty}^\infty d\tau \, e^{i\omega\tau} 
       \langle B_i^{I\dagger}(\tau)\, B_j^{I}(0)\rangle
\end{split}
\end{equation}
\begin{equation}
     \Lambda_{ij}(\omega) = -\frac{i}{2} \big(\Gamma_{ij}(\omega)-\Gamma_{ji}^*(\omega)\big)
\end{equation}
With these definitions, the master equation in the interaction picture becomes 
\begin{equation}
     \frac{d \rho_S^I(t)}{dt} = -i[H_{LS},\rho_S^I(t)] + \mathcal{D}[\rho_S^I(t)]
\end{equation}
The Lamb-shift Hamiltonian is $H_{LS}=\sum_{\omega} \sum_{ij} \Lambda_{ij} S_i^{\dagger}(\omega) S_j(\omega)$, 
and the dissipator is
\begin{equation*}
\begin{split}
    \mathcal{D}(\rho_S^I(t)) 
    &= \sum_{\omega}\sum_{i,j}G_{ij}(\omega) 
       \Big( S_j(\omega)\rho_S^I S_i^{\dagger}(\omega) \\
    &\quad - \tfrac{1}{2}\{ S_i^{\dagger}(\omega) S_j(\omega), \rho_S^I(t) \} \Big) 
\end{split}
\end{equation*}
Returning to the Schr\"odinger picture gives 
\begin{equation}
     \frac{d \rho_S(t)}{dt} = -i[H_S + H_{LS},\rho_S(t)] + \mathcal{D}[\rho_S(t)].
\end{equation}
The matrix $G_{ij}(\omega)$ is positive  for each transition frequency $\omega$. Therefore, it can be diagonalized to obtain Lindblad jump operators $L_\mu (\omega)$, which brings the master equation into the standard Lindblad form
\begin{equation}
\begin{split}
    \frac{d \rho_S(t)}{dt} 
    &= -i[H_S + H_{LS}, \rho_S(t)] \\
    &\quad \hspace{-1.0 cm}+ \sum_{\omega, \mu} \Big( L_\mu(\omega)\rho_S(t) L_\mu^{\dagger}(\omega) 
      - \tfrac{1}{2}\{ L_\mu^{\dagger}(\omega) L_\mu(\omega), \rho_S(t) \} \Big).
\end{split}
\end{equation}
The above equation is the Lindblad master equation. The Lamb-shift term $H_{\rm LS}$ represents a bath-induced correction to the system energy levels. When the primary focus is on relaxation and decoherence rates, this term can often be absorbed into the renormalized system Hamiltonian or neglected if it is small compared with the relevant transition energies.

\section{Analytical Solutions of the MQMF-LME}
\label{app:analytical}

\subsection{Noise-Free MQMF-LME}

The MQMF-LME for the system qubit can be written as 
\begin{align}
\frac{d \rho}{dt} =\;& -i [H, \rho] \nonumber \\
&+ \Gamma_{\mathrm{SE}} \left( \sigma_- \rho \sigma_+ - \frac{1}{2} \left\{ \sigma_+\sigma_-, \rho \right\} \right) \nonumber \\
&+ \Gamma_{\downarrow} \left( \sigma_- \rho \sigma_+ - \frac{1}{2} \left\{ \sigma_+ \sigma_-, \rho \right\} \right) \nonumber \\
&+ \Gamma_{\uparrow} \left( \sigma_+ \rho \sigma_- - \frac{1}{2} \left\{ \sigma_- \sigma_+, \rho \right\} \right),
\label{eq:appendix-complete-MQMF-LME}
\end{align}
Here, $H=(\omega_0/2)\sigma_z$, $\Gamma_{\mathrm{SE}}$ is the intrinsic spontaneous-emission rate, $\Gamma_{\downarrow}$ is the bath-induced downward transition rate, and $\Gamma_{\uparrow}$ is the bath-induced upward transition rate. Because $\Gamma_{\mathrm{SE}}$ and $\Gamma_{\downarrow}$ multiply the same lowering-channel dissipator, they can be combined into a single effective downward rate,
\[
\Gamma_{\text{SE}} + \Gamma_{\downarrow} = \Gamma_-,
\qquad
\Gamma_{\uparrow} = \Gamma_+ .
\]
With these definitions, Eq.~\ref{eq:appendix-complete-MQMF-LME} becomes
\begin{align}
\frac{d \rho}{dt} =\;& -i [H, \rho] + \Gamma_{-} \mathcal{D} [\sigma_{-}]\rho + \Gamma_{+} \mathcal{D[\sigma_{+}]\rho},
\end{align}
where $\mathcal{D} [L]\rho = L\rho L^{\dagger}-\frac{1}{2}\{L^{\dagger}L,\rho\}$. For a two-level system with density matrix
\begin{equation}
\rho = \begin{pmatrix}
\rho_{ee} & \rho_{eg} \\
\rho_{ge} & \rho_{gg},
\end{pmatrix}
\end{equation}
the Hamiltonian contribution is
\begin{equation}
-i[H,\rho]= \begin{pmatrix}
0 & -i\omega_0\rho_{eg} \\
+i\omega_0\rho_{ge} & 0
\end{pmatrix}.
\end{equation}
Now, the problem is much simplified,  to solve the LME, we need to focus on only two dissipation operations:
The effective downward dissipator is
\begin{equation}
    \Gamma_- \left( \sigma_- \rho \sigma_+ - \frac{1}{2} \left\{ \sigma_+\sigma_-, \rho \right\} \right) =
    \begin{pmatrix}
       -\Gamma_-\rho_{ee} & -\frac{\Gamma_-}{2}\rho_{eg}\\
-\frac{\Gamma_-}{2}\rho_{ge} & \Gamma_-\rho_{ee},
    \end{pmatrix}
    \label{eq:Matrix representation for relaxation dissipation}
\end{equation}
and the upward dissipator is
\begin{equation}
   \Gamma_{+} \left( \sigma_+ \rho \sigma_- - \frac{1}{2} \left\{ \sigma_- \sigma_+, \rho \right\} \right)=
    \begin{pmatrix}
       \Gamma_+\rho_{gg} & -\frac{\Gamma_+}{2}\rho_{eg}\\
-\frac{\Gamma_+}{2}\rho_{ge} & -\Gamma_+\rho_{gg}
    \end{pmatrix}
\label{eq: Matrix representation for excitation dissipation}
\end{equation}
Combining the Hamiltonian and dissipative contributions gives the coupled equations
\begin{equation}
   \dot\rho_{ee} = -\Gamma_-\rho_{ee}+\Gamma_+\rho_{gg}
   \label{eq: Differential euation for excited state population}
\end{equation}
\begin{equation}
   \dot\rho_{gg} = \Gamma_-\rho_{ee}-\Gamma_+\rho_{gg}
   \label{eq: Differential equation for ground state population}
\end{equation}
\begin{equation}
   \dot\rho_{eg} = -i\omega_0\rho_{eg}-\frac{\Gamma_+ + \Gamma_-}{2}\rho_{eg}
   \label{eq: Differential equation for off-diagonal elements 2}
\end{equation}
\begin{equation}
   \dot\rho_{ge} = i\omega_0\rho_{ge}-\frac{\Gamma_+ + \Gamma_-}{2}\rho_{ge}
   \label{eq: Differential equation for off-diagonal elements 1}
\end{equation}
Solving these first-order differential equations gives 
\begin{equation}
    \rho_{ee}(t) = \frac{\Gamma_+}{\Gamma_+ + \Gamma_-}+\left( \rho_{ee}(0)-\frac{\Gamma_+}{\Gamma_+ + \Gamma_-}\right)e^{-(\Gamma_+ +\Gamma_-)t}
    \label{eq: Solution1}
\end{equation}

\begin{equation}
    \rho_{gg}(t) = \frac{\Gamma_-}{\Gamma_+ + \Gamma_-}+\left( \rho_{gg}(0)-\frac{\Gamma_-}{\Gamma_+ + \Gamma_-}\right)e^{-(\Gamma_+ +\Gamma_-)t}
    \label{eq: Solution2}
\end{equation}

\begin{equation}
    \rho_{ge}(t) = \rho_{ge}(0) e^{i\omega_0t}e^{-\left({\frac{\Gamma_+ +\Gamma_-}{2}}\right)t}
    \label{eq: Solution3}
\end{equation}

\begin{equation}
    \rho_{eg}(t) = \rho_{eg}(0) e^{-i\omega_0t}e^{-\left({\frac{\Gamma_+ +\Gamma_-}{2}}\right)t}
    \label{eq: Solution4}
\end{equation}
These solutions show that the diagonal density-matrix elements relax toward finite steady-state populations, while the off-diagonal elements decay exponentially with rate $\frac{\Gamma_+ + \Gamma_-}{2}$. In the absence of an additional pure-dephasing channel, this gives the relaxation-limited relation $T_2 = 2T_1$.

\subsection{MQMF-LME with Noise-Induced Dephasing}

We next include noise-induced dephasing by allowing the qubit energy splitting to fluctuate in time. The Hamiltonian becomes
\begin{equation}
H(t)=\frac{\omega_0}{2}\sigma_z+\frac{f(t)}{2}\sigma_z = \frac{\omega_0+f(t)}{2}\sigma_z.
\end{equation}
where $f(t)$ is a stochastic fluctuation induced by the noisy environment. Defining
\begin{equation}
\Omega(t)=\omega_0+f(t),
\end{equation}
the MQMF-LME becomes
\begin{equation}
\frac{d\rho}{dt} = -i[H(t),\rho] + \Gamma_-\mathcal{D}[\sigma_-]\rho + \Gamma_+\mathcal{D}[\sigma_+]\rho.
\label{eq: Updated LME}
\end{equation}
The stochastic Hamiltonian contribution is
\begin{equation}
    -i[H(t),\rho] = 
    \begin{pmatrix}
        0 & -i\Omega(t)\rho_{eg}\\
        i\Omega(t)\rho_{ge} & 0
    \end{pmatrix}.
    \label{eq:Matrix representation for commutator}
\end{equation}
Because this term is diagonal in the energy eigenbasis, it does not directly change the populations.
\begin{align}
\dot{\rho}_{ee} &= -\Gamma_-\rho_{ee} + \Gamma_+\rho_{gg}, \label{eq:eom_rho_ee} \\
\dot{\rho}_{gg} &= \Gamma_-\rho_{ee} - \Gamma_+\rho_{gg}, \label{eq:eom_rho_gg} \\
\dot{\rho}_{eg} &= -\left[i\Omega(t) + \frac{\Gamma_-+\Gamma_+}{2}\right]\rho_{eg}, \label{eq:eom_rho_eg} \\
\dot{\rho}_{ge} &= \left[i\Omega(t) - \frac{\Gamma_-+\Gamma_+}{2}\right]\rho_{ge}. \label{eq:eom_rho_ge}
\end{align}
The diagonal equations are identical to the noise-free case, confirming that the stochastic term $f(t)\sigma_z/2$ produces pure dephasing without modifying the population dynamics. For the coherence $\rho_{eg}(t)$ , substituting $\omega_0 + f(t)$ gives 
\begin{equation}
\frac{d\rho_{eg}}{dt} = \left[-i\omega_0 - if(t) - \frac{\Gamma_-+\Gamma_+}{2}\right]\rho_{eg}.
\label{eq:coherence_diff}
\end{equation}
Solving this equation yields
\begin{equation}
\ln\left[\frac{\rho_{eg}(t)}{\rho_{eg}(0)}\right] = -i\omega_0 t - i\int_0^t f(t')dt' - \frac{\Gamma_-+\Gamma_+}{2}t.
\label{eq:solution of noise coherence}
\end{equation}
We define the accumulated stochastic phase as
\begin{equation}
\phi(t)=\int_0^t f(t')dt'.
\label{eq:stochastic_phase_def}
\end{equation}
Then
\begin{equation}
\rho_{eg}(t) = \rho_{eg}(0) e^{-i\omega_0 t} e^{-i\phi(t)} e^{-\frac{\Gamma_-+\Gamma_+}{2}t},
\label{eq:sol_rho_eg}
\end{equation}
and
\begin{equation}
\rho_{ge}(t) = \rho_{ge}(0) e^{i\omega_0 t} e^{i\phi(t)} e^{-\frac{\Gamma_-+\Gamma_+}{2}t}.
\label{eq:sol_rho_ge}
\end{equation}
To obtain the physical coherence decay, we average over noise realizations:
\begin{equation}
\langle \rho_{eg}(t)\rangle = \rho_{eg}(0) e^{-i\omega_0 t} e^{-\frac{\Gamma_-+\Gamma_+}{2}t} \left\langle e^{-i\phi(t)}\right\rangle,
\label{eq:avg_rho_eg_init}
\end{equation}
and 
\begin{equation}
\langle \rho_{ge}(t)\rangle = \rho_{ge}(0) e^{i\omega_0 t} e^{-\frac{\Gamma_-+\Gamma_+}{2}t} \left\langle e^{i\phi(t)}\right\rangle.
\label{eq:avg_rho_ge_init}
\end{equation}
We assume that $f(t)$ is a zero-mean Gaussian noise process. 
For any zero-mean Gaussian random variable $x$, its characteristic function satisfies~\cite{van1974cumulant}:
\begin{equation}
\left\langle e^{iax}\right\rangle = e^{-\frac{a^2}{2}\langle x^2\rangle}.
\label{eq:gaussian_identity}
\end{equation}
Using $x = \phi(t)$ and $a = \mp 1$, we obtain
\begin{equation}
\left\langle e^{-i\phi(t)}\right\rangle = \left\langle e^{i\phi(t)}\right\rangle = e^{-\frac{1}{2}\langle \phi^2(t)\rangle}.
\end{equation}
Defining the dephasing function
\begin{equation}
\chi(t) = \frac{1}{2}\langle \phi^2(t)\rangle,
\label{eq:chi_def}
\end{equation}
the phase average becomes
\begin{equation}
\left\langle e^{\pm i\phi(t)}\right\rangle = e^{-\chi(t)}.
\end{equation}
Since
\begin{equation}
\langle \phi^2(t)\rangle = \int_0^t dt_1 \int_0^t dt_2 \langle f(t_1)f(t_2)\rangle, 
\end{equation}
the dephasing function is
\begin{equation}
\chi(t) = \frac{1}{2} \int_0^t dt_1 \int_0^t dt_2 \langle f(t_1)f(t_2)\rangle.
\label{eq:chi_integral_form}
\end{equation}
Therefore, the ensemble-averaged coherences are
\begin{align}
\langle \rho_{eg}(t)\rangle &= \rho_{eg}(0) e^{-i\omega_0 t} e^{-\frac{\Gamma_-+\Gamma_+}{2}t} e^{-\chi(t)}, \label{eq:final_avg_rho_eg} \\
\langle \rho_{ge}(t)\rangle &= \rho_{ge}(0) e^{i\omega_0 t} e^{-\frac{\Gamma_-+\Gamma_+}{2}t} e^{-\chi(t)}. \label{eq:final_avg_rho_ge}
\end{align}

\subsection{Evolution Under Gaussian $1/f$ Noise}

We now consider Gaussian $1/f$ noise as a specific example of noise-induced pure dephasing. Starting from
\begin{equation}
\chi(t) = \frac{1}{2} \int_0^t dt_1 \int_0^t dt_2 \langle f(t_1)f(t_2)\rangle,
\end{equation}
we assume that the noise is stationary, so that
\begin{equation}
\langle f(t_1)f(t_2)\rangle = C_f(t_1-t_2).
\end{equation}
The autocorrelation function can be written in terms of the Fourier transform of the noise power spectral density $S_f(\omega)$ as
\begin{equation}
C_f(\tau) = \int_{-\infty}^{\infty} \frac{d\omega}{2\pi} S_f(\omega)e^{-i\omega \tau}.
\end{equation}
Substituting this expression into $\chi(t)$ gives 
\begin{equation}
    \chi(t) = \frac{1}{2} \int_{-\infty}^{\infty} \frac{d\omega}{2\pi} S_f(\omega) \left[ \int_0^t dt_1 e^{-i\omega t_1} \right] \left[ \int_0^t dt_2 e^{i\omega t_2} \right]
    \label{eq:Fourier Transform}
\end{equation}
The time integrals yield
\begin{equation}
\left| \int_0^t dt_1 e^{-i\omega t_1} \right|^2 = \frac{2[1-\cos(\omega t)]}{\omega^2}.
\end{equation}
Thus,
\begin{equation}
\begin{split}
\chi(t) &= \int_{-\infty}^{\infty} \frac{d\omega}{2\pi} S_f(\omega) \frac{1-\cos(\omega t)}{\omega^2}.
\end{split}
\end{equation}
For a classical real noise process, $S_f(-\omega) = S_f(\omega)$, and the expression can be written over positive frequencies as
\begin{equation}
\chi(t) = \frac{1}{\pi} \int_0^\infty d\omega\, S_f(\omega) \frac{1-\cos(\omega t)}{\omega^2}.
\label{eq:filterfunction_integral}
\end{equation}
For a two-sided $1/f$ noise spectrum, $S_f(\omega) = A/|\omega|$, where $A$ denotes the noise amplitude. Because the ideal $1/f$ diverges at both low and high frequencies, we introduce low-frequency and high-frequency cutoffs, $\omega_l$ and $\omega_h$.
The $1/f$ noise-dephasing function is therefore
\begin{equation}
\chi_{1/f}(t) = \frac{A}{\pi} \int_{\omega_l}^{\omega_h} d\omega\, \frac{1-\cos(\omega t)}{\omega^3}.
\end{equation}
Using the substitution $x = \omega t$, this becomes
\begin{equation}
\chi_{1/f}(t) = \frac{A t^2}{\pi} \int_{\omega_l t}^{\omega_h t} dx\, \frac{1-\cos x}{x^3}.
\label{eq:general_one_over_f_result}
\end{equation}
In the experimentally relevant regime
\begin{equation}
\omega_l t \ll 1 \quad \text{and} \quad \omega_h t \gg 1,
\end{equation}
the dominant contribution comes from the low-frequency part of the spectrum. Using $1-\cos x \approx x^2/2$ for small $x$, 
\begin{equation}
\int_{\omega_l t}^{\omega_h t} dx\, \frac{1-\cos x}{x^3} \approx \frac{1}{2} \int_{\omega_l t}^{1} \frac{dx}{x} = \frac{1}{2} \ln\left(\frac{1}{\omega_l t}\right).
\end{equation}
Thus,
\[
\chi_{1/f}(t) \approx \frac{A t^2}{2\pi} \ln\left(\frac{1}{\omega_l t}\right).
\]
Substituting this result into the ensemble-averaged coherence gives
\begin{equation}
\langle \rho_{eg}(t)\rangle = \rho_{eg}(0) e^{-i\omega_0 t} e^{-\frac{\Gamma_++\Gamma_-}{2}t} \exp\left[ -\frac{A t^2}{2\pi} \ln\left(\frac{1}{\omega_l t}\right) \right].
\end{equation}
The corresponding coherence magnitude is
\begin{equation}
|\langle \rho_{eg}(t)\rangle| = |\rho_{eg}(0)| e^{-\frac{\Gamma_++\Gamma_-}{2}t} \exp\left[ -\frac{A t^2}{2\pi} \ln\left(\frac{1}{\omega_l t}\right) \right].
\end{equation}
This expression shows that Gaussian $1/f$ noise produces a non-exponential dephasing envelope in addition to the exponential coherence decay from the Lindblad relaxation channels. The first term in the exponent represents relaxation-induced coherence decay, while the second term represents the additional pure-dephasing contribution from $1/f$ noise. Because the $1/f$-noise contribution is not linear in time, the total coherence decay is generally non-exponential. 
An effective decoherence time $T_2$ can be obtained by fitting the magnitude of the ensemble-averaged coherence, $|\langle \rho_{eg}(t)\rangle|$ to a single exponential decay over the same time window used for the noise-free simulations.

\section{Estimation of the Characteristic FRET Distance}
\label{app:fret-distance}

In the conventional FRET description, the characteristic F\"orster distance $R_0$ denotes the donor--acceptor separation at which the excitation-transfer efficiency reaches 50\%. In the MQMF-LME framework, $R_0$ enters the distance-dependent transfer rate through Eq.~\ref{eq:gamma-FRET}, where it sets the effective strength of the FRET-mediated interaction. Larger values of $R_0$ correspond to stronger excitation transfer over a given donor--acceptor separation. As shown in Eq.~\ref{eq:critical transfer}, $R_0$ depends on the donor quantum yield $Q_D$ and the spectral overlap integral $J$, while parameters such as the dipole-orientation factor \(\kappa\) and refractive index \(n\) are fixed once the host material and orientational averaging assumptions are specified. For the Er$^{3+}$–Er$^{3+}$ FRET interaction, $Q_D$ denotes the quantum yield of the donor Er$^{3+}$ ion in the absence of an acceptor Er$^{3+}$ ion. A larger $Q_D$ indicates fewer competing nonradiative decay pathways and a longer effective donor excited-state lifetime, thereby increasing the probability that excitation transfer can occur through the FRET channel. The spectral overlap integral $J$ quantifies the overlap between the donor emission spectrum and the acceptor absorption spectrum. Equations~\ref{eq:Spectral overlap integral} and~\ref{eq:Spectral Overlap integral calculation} define how $J$ is estimated from the absorption cross section $\sigma$. A larger absorption cross section increases the molar absorption coefficient and therefore increases $J$, which in turn increases $R_0$ and enhances the effective donor--acceptor excitation-transfer rate.

Both $Q_D$ and $\sigma$ are difficult to determine precisely for Er$^{3+}$-doped CeO$_2$, because they depend strongly on the local environment of the Er$^{3+}$ ions. Local crystal-field distortions, localized defect states, and impurity-induced polarization can modify transition energies and oscillator strengths, while evanescent and reactive near-field effects can modify donor--acceptor coupling. These effects make $Q_D$ and $\sigma$ physically uncertain even when other parameters, such as $\kappa$ and $n$, are held fixed. To quantify this uncertainty, we vary $Q_D$ and $\sigma$ and calculate the corresponding values of $R_0$, as shown in Fig.~\ref{fig:Phase-diagram}. The white contour lines in the phase diagram indicate representative $R_0$ values ranging from 0.8 to 2.5 nm. Higher values of $Q_D$ and $\sigma$ increase $R_0$, reflecting fewer competing nonradiative decay pathways and stronger donor--acceptor spectral overlap. Conversely, lower values of $\sigma$ reduce the spectral overlap and decrease the effective FRET distance. Across the physically reasonable range $R_0$ = 0.8--2.5 \text{nm}, the MQMF-LME simulations show the closest agreement with experiment at  $R_0= 0.98$ nm. Importantly, this value is not associated with a unique pair of $Q_D$ and $\sigma$. The same effective $R_0$ can arise from different combinations of donor quantum yield and absorption cross section, indicating that $R_0$ should be interpreted as an effective interaction parameter that incorporates multiple local-environment effects.

In Fig.~\ref{fig:Phase-diagram}, the contour corresponding to $R_0 = 0.98$ nm can be realized through three representative parameter regimes, labeled I, II, and III. In Region I, $Q_D$ is high while $\sigma$ is low, indicating that the donor has few competing nonradiative decay pathways but weak acceptor absorption at the donor emission wavelength. In Region II, both $Q_D$ and $\sigma$ take moderate values. In Region III, $Q_D$ is lower because of stronger nonradiative decay, but $\sigma$ is larger, indicating stronger acceptor absorption and larger spectral overlap. This non-unique mapping between $Q_D$, $\sigma$, and $R_0$ is physically important. It shows that the best-fit value $R_0=0.98$ nm should not be interpreted as a uniquely determined microscopic constant. Rather, it is an effective FRET interaction length that captures the combined influence of donor quantum yield, acceptor absorption strength, spectral overlap, and local-environment effects. The fact that $R_0
=0.98$ nm lies within the physically plausible region of the phase diagram supports the use of FRET-mediated excitation transfer as the dominant mechanism for the concentration-dependent relaxation observed in Er$^{3+}$-doped CeO$_2$.

\bibliography{lit}
\end{document}